\documentclass[letterpaper]{article}

\usepackage{graphicx,lscape}
\usepackage{amsmath,amssymb}
\usepackage{bm}
\usepackage{mathrsfs}
\usepackage{url}
\usepackage{natbib}

\usepackage{amsmath,amssymb}
\usepackage{bm}
\usepackage[figuresright]{rotating}
\usepackage{color}
\usepackage{caption}
\usepackage {lscape}
\usepackage{ulem}

\begin{document}

\begin{center}
\Large \bf On a fundamental problem in the analysis of cancer registry data
\end{center}

\vspace{3mm}

\begin{center}
Sho Komukai \\
{\it Department of Biomedical Statistics, Graduate School of Medicine, Osaka University\\
Yamadaoka 2-2, Suita City, Osaka 565-0871, Japan \\
E-mail: skomukai@biostat.med.osaka-u.ac.jp
} \\ 
\vspace{2mm}
Satoshi Hattori \\
{\it Department of Biomedical Statistics, Graduate School of Medicine, \\ and \\
Institute for Open and Transdisciplinary Research Initiatives, Osaka University\\
Yamadaoka 2-2, Suita City, Osaka 565-0871, Japan \\
E-mail: hattoris@biostat.med.osaka-u.ac.jp
} \\ 
\vspace{2mm}
and \\
\vspace{2mm}
Bernard Rachet \\
{\it Inequalities in Cancer Outcomes Network, Department of Non-communicable Disease Epidemiology, Faculty of Epidemiology and Population Health, London School of Hygiene and Tropical Medicine. \\
Keppel Street, London WC1E 7HT, UK \\
E-mail: Bernard.Rachet@lshtm.ac.uk
} \\ 
\end{center}

\begin{center}
March 16, 2023
\end{center}

\vspace{1mm}

\begin{abstract}
In epidemiology research with cancer registry data, it is often of primary interest to make inference on cancer death, not overall survival. 
Since cause of death is not easy to collect or is not necessarily reliable in cancer registries, some special methodologies have been introduced and widely used by using the concepts of the relative survival ratio and the net survival. 
In making inference of those measures, external life tables of the general population are utilized to adjust the impact of non-cancer death on overall survival. 
The validity of this adjustment relies on the assumption that mortality in the external life table approximates non-cancer mortality of cancer patients. 
However, the population used to calculate a life table may include cancer death and cancer patients. 
Sensitivity analysis proposed by Talb\"{a}ck and Dickman to address it requires additional information which is often not easily available. 
We propose a method to make inference on the net survival accounting for potential presence of cancer patients and cancer death in the life table for the general population. 
The idea of adjustment is to consider correspondence of cancer mortality in the life table and that in the cancer registry. 
We realize a novel method to adjust cancer mortality in the cancer registry without any additional information to the standard analyses of cancer registries. 
Our simulation study revealed that the proposed method successfully removed the bias. 
We illustrate the proposed method with the cancer registry data in England. 
\end{abstract}

\begin{center}
\end{center}

\newpage
\baselineskip=24pt

\section{Introduction}
Cancer registries provide comprehensive and useful information on cancer and are utilized to conduct various epidemiology research including nation-wide comparisons of cancer survival and estimation of change in cancer survival. \cite{Angelis2014, Allemani2018} 
Cancer survival is an important measure. 
However, collecting reliable and consistent information on the cause of death is challenging. 
To make inference on survival from cancer without relying on the cause of death information (i.e., within the relative survival data setting), special survival analysis techniques have been then developed and widely used in analyses of cancer registry data. 
Several cancer survival measures using such techniques help to describe the survival experience of cancer patients, including relative survival ratio, net survival, or crude probabilities of death. \cite{Ederer1961, Cronin2000, Perme2012, Perme2016, Belot2019} 
The techniques to estimate such measures are based on the assumption that the hazard of death can be decomposed into the hazard of death due to the cancer of interest and that due to other causes. 
In the absence of reliable information on the cause of death, as proposed in contexts other than cancer registries \cite{Breslow1983}, an option is to borrow external information, in order to estimate the hazard of death due to other causes from the general population to which the patient belongs. 
Mortality hazards for the general population are available in population life tables based on demographic statistics, which are published in most countries at least by age, sex and calendar period. 

However, it assumes that the cancer deaths contained in these population life tables are too few to affect the estimation of the mortality hazard due to other causes. 
This assumption which may not always hold has been discussed by a few authors. \cite{Ederer1961, Esteve1994, Talback2011} 
\citet{Ederer1961} claimed that since the sizes of age-, gender- and site-specific subpopulations of cancer patients was much smaller than their counterparts in the general population, the impact of cancer deaths contained in the general population was negligible. 
However, it may not be true for all cancer types and subpopulations. 
\citet{Talback2011} utilized uncommon life tables, which contained individual information on cancer patients included in these life tables. 
Considering the date of cancer diagnosis as censoring, they were able to estimate the non-cancer mortality hazard in the general population. 
They concluded that the presence of cancer deaths in the life tables hardly impacted the cancer survival estimation in most situations, but they also observed some bias in a few subpopulations. 
Given the general unavailability of such life tables (i.e., with individual information), \citet{Talback2011} also introduced a model-based method to conduct a sensitivity analysis without such individual-level information. 
However, their model for sensitivity analysis required the number of cancer deaths in the general population and did not account for inclusion of cancer patients in the general population.
Since the number of cancer deaths cannot be obtained from cancer registries or is not reliably known in this setting, the model-based sensitivity analysis method by \citet{Talback2011} might not be easily applicable. 

In this paper, we propose a method to estimate cancer survival measures in the relative survival setting, while accounting for the potential enrollment of cancer patients and cancer deaths in the life tables. 
To this end, we only rely on additional information on cancer incidence, which is usually publicly available. 
Even if unavailable in public, incidence rates can be calculated with vital statistics coupled with the cancer registry. 
Thus, we do not need to take much efforts to gather additional information to apply our method and then eliminate biases due to cancer deaths in the life table. 
The key idea of our development is to adjust survival of cancer patients in the life-table by borrowing information from the cancer registry.
We illustrate the method with the most used cancer survival measure, net survival, which is the survival probability of cancer patients in the hypothetical situation of individuals who can only die from their cancer. 
More specifically, we describe the method in the application to a non-parametric estimator of the net survival called Pohar-Perme estimator. \cite{Perme2012} 

The organization of the rest of this paper is as follows. 
In Section 2.1, we introduce the net survival measure in the relative survival setting and its Pohar-Perme estimator. 
In Section 2.2, we discuss the assumptions implicitly made when using life tables in the relative survival setting. 
Section 3 presents the notations of the quantities, which are used in our approach, from the life tables. 
Section 4 details the key components of our approach: in Section 4.1, we introduce the incidence rate used in our method and explain how to estimate it from cancer registry data if unavailable in public; in Section 4.2, we summarize the assumptions to utilize the correspondence between cancer registry and life tables. 
In Section 4.3, we introduce an integral equation to obtain the non-cancer survival for the general population with adjustment for the cancer deaths.
In Section 4.4, we show how to solve an empirical version of the integral equation.
We evaluate the proposed method by the simulation studies in Section 5 and illustrate it on real cancer registry data in England in Section 6.
Conclusions and some discussions are given in Section 7.
Detailed formulas and all theoretical details are given in Appendixes.

\section{Estimation of the net survival}
\label{sec_est_net}
\subsection{Pohar-Perme estimator}
\label{sec_PP}
Let $Z_D$ be a vector of the baseline covariates recorded in the cancer registry such as age at diagnosis, year of diagnosis, gender and cancer stage at diagnosis. 
The subscript ``$D$" is attached to covariates that are observed at the time of cancer diagnosis.
Denote the time from diagnosis to death due to any causes by $T_O$. 
We assume that $T_O$ can be right-censored by the potential censoring time $C$. 
Thus, the observed components are $T=\min(T_O,C)$, the indicator of censoring $\Delta=I(T_O\le C)$, and the covariates $Z_D$, where $I(\cdot)$ is the indicator function. 
Let $T_E$ and $T_P$ be the time to death due to cancer and that due to any causes other than cancer, respectively, from the date at cancer diagnosis. 
Note that $T_O=\min(T_E,T_P)$. We assume that $T_E$ and $T_P$ are continuous. 
We suppose we observe $n$ i.i.d. copies of $(T,\Delta,Z_D)$, and $(T_i,\Delta_i,Z_{D,i})$ is the observation for the $i$th subject $(i=1, 2, ..., n)$. 
We make inference based on these observations. 
For any random variable, we use the subscript $i$ for representing its counterpart for the $i$th subject. 

The survival function for $T_O$ is denoted by $S_O(t)= \Pr(T_O>t)$ and the corresponding hazard and cumulative hazard functions are denoted by $\lambda_O(t)$ and $\Lambda_O(t)$, respectively. 
The survival, hazard and cumulative hazard functions conditional on $Z_D$ are denoted by $S_O(t|Z_D)$, $\lambda_O(t|Z_D)$ and $\Lambda_O(t|Z_D)$, respectively. 
The corresponding quantities for $T_E$, $T_P$, and $C$ are denoted in a similar way with the subscript ``$E$", ``$P$", and ``$C$", respectively. 

The net survival is defined as $S_E(t)=P(T_E>t)$, which is the marginal survival function of $T_E$ and the estimand of interest. 
The $PP$ estimator is defined by 
\begin{eqnarray}
\hat{\Lambda}_E^{PP} (t)
= \int _{0}^{t} \frac{\sum _{i=1}^{n} \frac{1 }{S_P(u|Z_{D,i}) }\left\{dN_i (u) -Y_i(u) d\Lambda_P (u|Z_{D,i})\right\}}{\sum _{j=1}^{n} \frac{Y_j (u) }{S_P(u|Z_{D,j}) } }. 
\label{PP}
\end{eqnarray}
where $N_i(t)$ and $Y_i(t)$ are the counterpart for the $i$th subject of the counting process $N(t)=I(T \le t, \Delta=1)$ and at-risk process $Y(t)=I(T>t)$, respectively.\cite{Perme2012} 
Assuming that $S_P(t|Z_D)$ and $\Lambda_P(t|Z_D)$ are known for any $Z_D$ and $t$, \citet{Perme2012} showed that the $PP$ estimator consistently estimates $\Lambda_E(t)$ under the conditions $({\rm A\mathchar`-1}) \ T_E\perp T_P|Z_D$ and $({\rm A\mathchar`-2})\ C\perp \{T_E,T_P, Z_D\}$ (independent censoring). 

\subsection{Extracting $S_P(t|Z_D)$ from a life table}
\label{S_P_extracting}
In practice, as $S_P(t|Z_D)$ is unknown, a life table for the general population is used to calculate $S_P(t|Z_D)$ by extracting the survival function of the general population with the same covariates from the life table. 
The information contained in the life table is defined by some socio-demographic variables such as age, calendar year, and gender. 
Suppose that $Z_D$ has no cancer-specific variable. 
More specifically, let $Z_D=(Z_D^{(1)},Z_D^{(2)},Z_D^{(others)tr})^{tr}$, where $Z_D^{(1)}$ and $Z_D^{(2)}$ are the age at cancer diagnosis and the year of cancer diagnosis, respectively, both being time-dependent variables, $Z_D^{(others)}$ be a column vector of other time-invariant demographic covariates, such as gender and race, and for any column vector $V$, $V^{tr}$ indicates the transpose of $V$. 

From the definition, $S_P(t|Z_D)$ is the survival function for $T_P$, which is the time to non-cancer death if the subject would not die from cancer since the date of their cancer diagnosis. 
Although extracting the survival function corresponding to $S_P(t|Z_D)$ from life tables is a widely-used practice, we would like to discuss its appropriateness more carefully. 
Life tables provide annual mortality rates for the population of specific age, calendar year, and $Z_D^{(others)}$. 
With a series of life tables, we can construct a lexis diagram as shown in Figure~\ref{f:relation}($b$). 
Note that we consider $Z_D^{(others)}$ fixed to a single value since this Lexis diagram is created for each value of $Z_D^{(others)}$.
For cancer patients of interest with covariates $Z_D$ (say, 50 years old in 1990, as seen in Figure~\ref{f:relation}($a$)), the corresponding life table (matching on both age and year) is presented by the plain circle in Figure~\ref{f:relation}($b$). 
The survival function of the population by age and calendar year can be then extracted from the series of life tables on the diagonal line. 
We pretend there is a cohort of the population with this survival function. 
The validity of extracting $S_P(t|Z_D)$ with this survival function from the life-table is justified if 
\begin{description}
\item[(i)] No cancer patients are included in the cohort underlying the extracted survival function and the non-cancer subjects in this cohort do not die from cancer. 
\item[(ii)] The survival function for the time to non-cancer death of the cancer patients included in the cancer registry data is the same to the survival function from the life table, given the same background covariates. 
\end{description}
Even supposing the assumption (ii), the assumption (i) may be questionable in reality; some cancer patients can be included in the cohort underlying the extracted survival function, and they are more likely to die of cancer, whereas some non-cancer subjects can be diagnosed with cancer after being included in that cohort and can die of cancer. 

\section{Formulating the life table and revisiting the current practice}
\label{sec_f_life_table}
\begin{figure}
\centering\includegraphics[width=150mm,height=180mm]{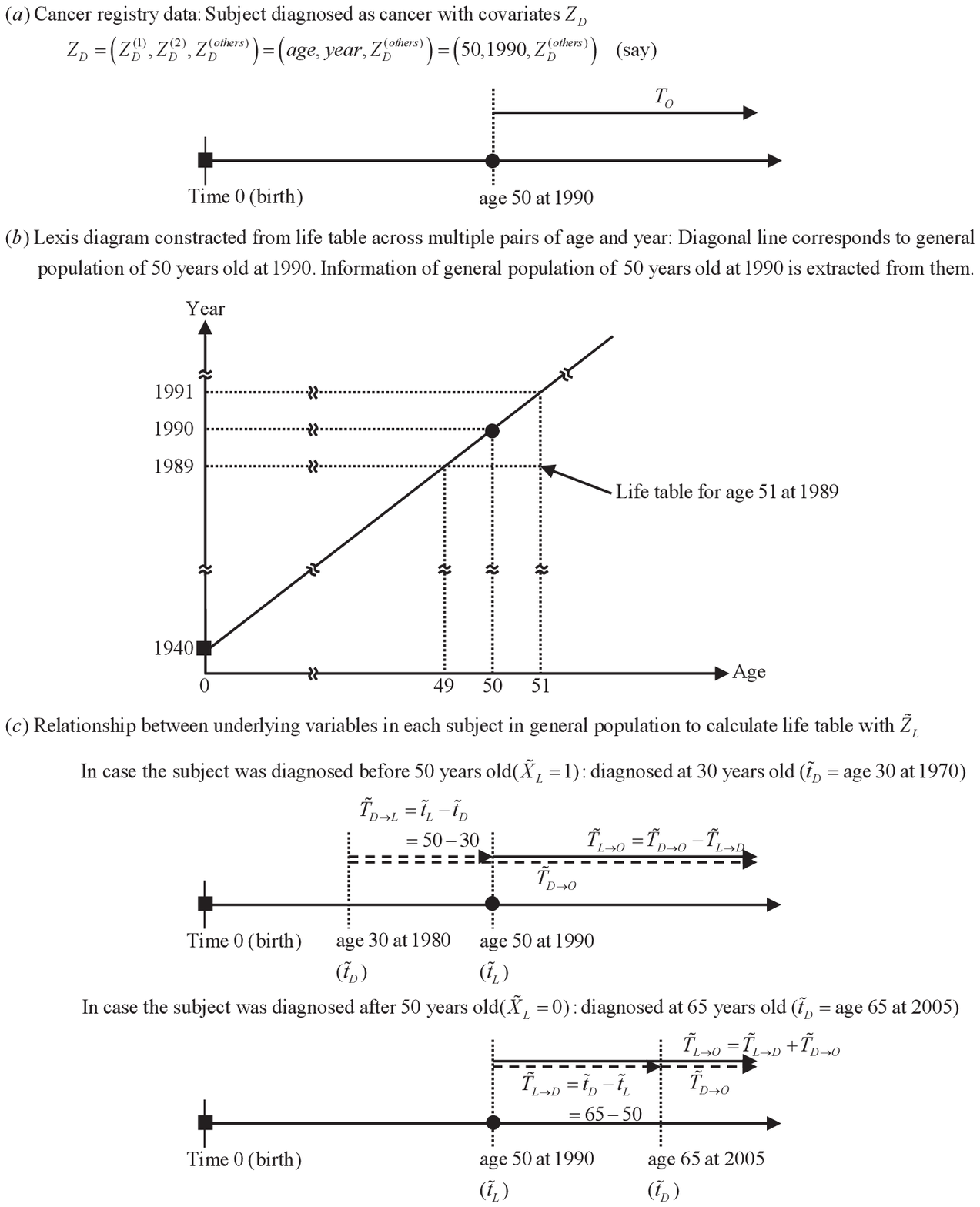}
\caption{Conceptual diagrams of cancer registry data and of corresponding life tables: (a) Line plot in age for a cancer patient with covariates at diagnosis, $Z_D$, in cancer registry data. (b) Lexis diagram constructed from life table across the pairs of age and year: for a cancer patient described in (a), information of the general population is extracted from them on diagonal line. (c) The relationship among the random variables for a subject registered in the life table.}
\label{f:relation}
\end{figure}

In this section, notations of the random variables related to the life tables are introduced, because we distinguish them from the notations applying to random variables related to the cancer registry.
For the cancer registry, we use the notations introduced in Section~\ref{sec_est_net}. 
For the life table, a tilde is systematically added. 
Let $\tilde{Z}_L=(\tilde{Z}_L^{(1)}, \tilde{Z}_L^{(2)}, \tilde{Z}_L^{(others)tr})^{tr}$ be a vector of covariates in the life table. 
Note that $\tilde{Z}_L$ has the same components as $Z_D$ and both of $Z_D$ and $\tilde{Z}_L$ vary only yearly. 
Suppose we consider a patient in the cancer registry, for example, who is diagnosed at age 50 in 1990. 
That is, $Z_D^{(1)}=50$ and $Z_D^{(2)}=1990$. See Figure~\ref{f:relation}($a$). 
The patient is matched with the life table of the corresponding covariates $(\tilde{Z}_L^{(1)}, \tilde{Z}_L^{(2)}, \tilde{Z}_L^{(others)tr})=(50, 1990, Z_D^{(others)tr})$, which is represented by a plain circle in Figures~\ref{f:relation}($a$) and \ref{f:relation}($b$). 
As mentioned in Subsection~\ref{S_P_extracting}, the corresponding survival function can be extracted from a series of life tables on the diagonal line through this plain circle, assuming the existence of a cohort underlying this survival function. 
This is illustrated with two specific subjects of this cohort in Figure~\ref{f:relation}($c$); one had been diagnosed as a cancer at the age of 50 ($\tilde{X}_L=1$; the upper panel of Figure~\ref{f:relation}($c$)) and the other had not ($\tilde{X}_L=0$; the lower panel of Figure~\ref{f:relation}($c$). 
To describe these individuals, we introduce $\tilde{t}_D$, which is the age at diagnosis. 
Define $\tilde{Z}_{D}=(\tilde{Z}_D^{(1)}, \tilde{Z}_D^{(2)}, \tilde{Z}_D^{(others)tr})^{tr}$ be a covariate vector at $\tilde{t}_D$. 
For notational convenience, set $\tilde{t}_L=\tilde{Z}_L^{(1)}$ at $\tilde{Z}_L^{(2)}$ ($\tilde{Z}_L^{(2)}=1990$ for the above illustrative patient). 
Let $\tilde{X}_L$ be a binary random variable, with the value 1 if, at $\tilde{t}_L$, subject had already been diagnosed with a cancer and the value 0 otherwise. 
To link $\tilde{Z}_L$ and $\tilde{Z}_D$, we use the notation $\tilde{Z}_{L\pm s}$ representing $\tilde{Z}_L$ after/before $s$ years from $\tilde{t}_L$. 
That is, $\tilde{Z}_{L\pm s}=(\tilde{Z}_L^{(1)}\pm s, \tilde{Z}_L^{(2)} \pm s, \tilde{Z}_L^{(others)tr})^{tr}=(age\pm s, year\pm s, \tilde{Z}_L^{(others)tr})^{tr}$. 
If a subject with $\tilde{Z}_L$ is diagnosed with a cancer after/before $s$ years from $\tilde{t}_L$, $\tilde{t}_D=\tilde{t}_L\pm s$, then $\tilde{Z}_D=\tilde{Z}_{L \pm s}$ holds. 
Similarly, we define $\tilde{X}_{L \pm s}$ as the information on $\tilde{X}_L$ at the time of $\tilde{t}_L \pm s$.

Let $\tilde{T}_{L \to O}$ be the time to death due to any cause from the date of $\tilde{t}_L$. 
In the subscript ``$L \to O$", ``$L$" in the left-hand side of the arrow means the origin and the right component corresponds to the event. 
Thus $\tilde{T}_{L \to E}$ and $\tilde{T}_{L \to P}$ are defined as the time-to-death due to cancer and non-cancer causes, respectively, from the date of $\tilde{t}_L$. 
The random variables for the time-to-death from cancer diagnosis such as $\tilde{T}_{D \to O}$, $\tilde{T}_{D \to E}$ and $\tilde{T}_{D\to P}$ are defined in a similar way. 
We also consider a random variable $\tilde{T}_{D \to L}= \tilde{t}_L - \tilde{t}_D$, which the time elapsed between the cancer diagnosis and $\tilde{t}_L$ for individuals whose cancer was diagnosed before the date $\tilde{t}_L$ (that is, $\tilde{X}_L=1$). 
For the non-cancer subjects, the corresponding random variable at the date of registration into the life table (i.e. $\tilde{X}_L=0$), is defined in the same way $\tilde{T}_{L \to D}=\tilde{t}_D-\tilde{t}_L$.

Let $\alpha(\tilde{z}_L)=\Pr(\tilde{X}_L=1|\tilde{Z}_L=\tilde{z}_L)$. 
Denote $\tilde{S}_{L \to O}(t|\tilde{Z}_L=\tilde{z}_L)=P(\tilde{T}_{L \to O}>t|\tilde{Z}_L=\tilde{z}_L)$. 
The corresponding survival functions for $\tilde{T}_{L \to E}$ and $\tilde{T}_{L \to P}$ are denoted in a similar way with the subscript ``$L \to E$" and ``$L \to P$", respectively. 
Let $\tilde{F}_{D \to L}(t|\tilde{Z}_L=\tilde{z}_L,\tilde{X}_L=1)=\Pr(\tilde{T}_{D\to L} \le t|\tilde{Z}_L=\tilde{z}_L,\tilde{X}_L=1)$ and $\tilde{F}_{L \to D}(t|\tilde{Z}_L=\tilde{z}_L,\tilde{X}_L=0)=\Pr(\tilde{T}_{L\to D} \le t|\tilde{Z}_L=\tilde{z}_L,\tilde{X}_L=0)$. 

In the current practice, $S_P(t|Z_D=z)$ is extracted with $\tilde{S}_{L \to O}(t|\tilde{Z}_L=z)$ matching $Z_D=\tilde{Z}_L$. If $\alpha(\tilde{z}_L)=0$ for any $\tilde{z}_L$ (no cancer patients are included in the general population used for the life table) and $\tilde{S}_{L \to E}(t|\tilde{Z}_L=\tilde{z}_L, \tilde{X}_L=0)=1$ for all $\tilde{z}_L$ (non-cancer subjects included in the life table do not die of cancer), $\tilde{S}_{L \to O}(t|\tilde{Z}_L=\tilde{z}_L, \tilde{X}_L=0)=\tilde{S}_{L \to P}(t|\tilde{Z}_L=\tilde{z}_L, \tilde{X}_L=0)$. 
Then, the assumption (i) in Section~\ref{sec_est_net} holds. The assumption (ii) in Section~\ref{sec_est_net} can be described as $\tilde{S}_{L \to P}(t|\tilde{Z}_L=\tilde{z}_L, \tilde{X}_L=0)=\tilde{S}_{L\to P}(t|\tilde{Z}_L=\tilde{z}_L)=S_{P}(t|Z_D=\tilde{z}_L)$ under the assumption (i). 
Then, the current practice is justified.

\section{Estimation of the net survival in the presence of cancer death in the life table}
\label{main_sec}
\subsection{Incidence rate}
\label{inc}
As described in Section~\ref{sec_est_net}, the standard analysis of cancer registry data requires the cancer registry data and the life table. 
In addition to these two datasets, we suppose that the information on the annual cancer incidence rate for each $\tilde{Z}_L$ is available. 
Let $f_{\tilde{t}_D}(u|\tilde{Z}_L=\tilde{z}_L, \tilde{X}_L=0)$ be the probability density function of $\tilde{t}_D$ conditional on $\tilde{Z}_L=\tilde{z}_L$ and $\tilde{X}_L=0$. 
The annual cancer incidence rate for the population with $\tilde{Z}_L=\tilde{z}_L$ is defined by
\begin{align}
IR(\tilde{z}_L)=\int_{\tilde{t}_L}^{\tilde{t}_L+1}{f_{\tilde{t}_D}(u|\tilde{Z}_L=\tilde{z}_L, \tilde{X}_L=0)du}. 
\label{def_IR}
\end{align}
In practice, $IR(\tilde{z}_L)$ are calculated as the number of new cancer patients diagnosed within a year divided by the number of person-years (within a year) in the general population with $\tilde{Z}_L=\tilde{z}_L$. 
The number of new cancer patients is calculated from the cancer registry, and the number of person-years (within a year) in the general population is calculated from the vital statistics. 
Thus the $IR(\tilde{z}_L)$ for each cancer type, even if unavailable in public, can be calculated from the cancer registry data and the vital statistics. 

\subsection{Assumptions}
\label{assum}
Although some assumptions were mentioned in the previous sections, we summarize all the assumptions for quantities in the cancer registry, those in the life table, and the relationship among them. Let
\begin{description}
\item[(A-1)] $T_E \perp T_P|Z_D$
\item[(A-2)] $C \perp \{T_E, T_P, Z_D\}$
\item[(B-1)] $\tilde{T}_{L \to E} \perp \tilde{T}_{L \to P}|\{ \tilde{Z}_L, \tilde{X}_L=0 \}$
\item[(B-2)] $\tilde{T}_{L \to D} \perp \tilde{T}_{D \to E}|\{ \tilde{Z}_{L}, \tilde{X}_L=0 \}$
\item[(B-3)] $\tilde{T}_{D \to L} \perp \tilde{T}_{L \to O}|\{ \tilde{Z}_{L}, \tilde{X}_L=1 \}$
\item[(C-1)] $S_E(t|Z_D=\tilde{z})=\tilde{S}_{D \to E}(t|\tilde{Z}_D=\tilde{z}, \tilde{X}_L=0)$
\item[(C-2)] $S_P(t|Z_D=\tilde{z})=\tilde{S}_{L \to P}(t|\tilde{Z}_{L}=\tilde{z}, \tilde{X}_L=0)$. 
\item[(C-3)] $S_O(t|Z_D=\tilde{z})=\tilde{S}_{D \to O}(t|\tilde{Z}_D=\tilde{z})=\tilde{S}_{D \to O}(t|\tilde{Z}_D=\tilde{z}, \tilde{X}_L=1)$
\end{description}

The assumptions (A-1) and (A-2) apply to the cancer registry data and are required by Poher-Perme estimator (Perme, Stare, and Est\`eve, 2012). 
The assumptions (B-1) to (B-3) apply to the life table. 
The assumption (B-1) corresponds to (A-1). 
As argued in Section~\ref{sec_f_life_table}, each cancer patient in the cancer registry is matched with a subject with the corresponding baseline characteristics in the cohort underlying in the life table. 
The corresponding survival function is then extracted (see Section~\ref{sec_f_life_table} and Figure~\ref{f:relation}). 
Assumptions (B-2) and (B-3) describe a kind of non-informativeness for extracted $T_P$ and $T_E$; once the baseline characteristics are matched, a subject in the life table is selected regardless of their natural history of cancer. 
The assumptions from (C-1) to (C-3) establish the correspondences between the cancer registry and life table data. 
The assumption (C-1) implies that the survival functions of the time to cancer death from diagnosis are the same between the cancer patients registered in the cancer registry and those in the life table if they have the same covariates at diagnosis. 
The assumption (C-2) means that the survival functions of the time to non-cancer death are common among the cancer patients and the non-cancer subjects as long as the baseline covariates are same. 
The assumption (C-2) guarantees assumption (ii) of Section~\ref{sec_est_net}.
The assumption (C-3) implies that cancer patients included in the life table are assumed to be similar to those in the cancer registry once diagnosed as cancer. 

\subsection{Integral equation for $S_P(t|Z_D)$}
\label{subsec_est_SP}
Recall that $\alpha(\tilde{z}_L)=\Pr(\tilde{X}_L=1|\tilde{Z}_L=\tilde{z}_L)$. It holds that
\begin{align}
\tilde{S}_{L \to O}(t|\tilde{Z}_L=\tilde{z}_L) 
&= \alpha(\tilde{z}_L)\tilde{S}_{L \to O}(t|\tilde{Z}_L=\tilde{z}_L, \tilde{X}_L=1) \nonumber \\
&+ \{1-\alpha(\tilde{z}_L)\}\tilde{S}_{L \to O}(t|\tilde{Z}_L=\tilde{z}_L, \tilde{X}_L=0). 
\label{eq1}
\end{align}
From (B-1), $\tilde{S}_{L\to O}(t|\tilde{Z}_L=\tilde{z}_L, \tilde{X}_L=0)=\tilde{S}_{L\to E}(t|\tilde{Z}_L=\tilde{z}_L, \tilde{X}_L=0) \times \tilde{S}_{L\to P}(t|\tilde{Z}_L=\tilde{z}_L, \tilde{X}_L=0)$. 
Then, by simple algebraic manipulation, the equation ($\ref{eq1}$) leads to
\begin{align}
\frac{\tilde{S}_{L \to O}(t|\tilde{Z}_L=\tilde{z}_L) - \alpha(\tilde{z}_L)\tilde{S}_{L \to O}(t|\tilde{Z}_L=\tilde{z}_L, \tilde{X}_L=1)}{\{1-\alpha(\tilde{z}_L)\}\tilde{S}_{L\to P}(t|\tilde{Z}_L=\tilde{z}_L, \tilde{X}_L=0)}=
\tilde{S}_{L \to E}(t|\tilde{Z}_L=\tilde{z}_L, \tilde{X}_L=0).
\label{eq2}
\end{align}
Under the assumption (C-2), $\tilde{S}_{L\to P}(t|\tilde{Z}_L=\tilde{z}_L, \tilde{X}_L=0)=S_P(t|Z_D=\tilde{z}_L)$. 
As presented in Appendix A, it holds that
\begin{align}
& \tilde{S}_{L \to E}(t|\tilde{Z}_L=\tilde{z}_L, \tilde{X}_L=0) \nonumber \\
&= 1 - \int_{0}^{t} \left\{ 1 - 
\frac{S_{O}(t-s|Z_D=\tilde{z}_{L+s})}{S_{P}(t-s|Z_D=\tilde{z}_{L+s})}
\right\} d\tilde{F}_{L\to D}(s|\tilde{Z}_L=\tilde{z}_{L}, \tilde{X}_L=0).
\label{s_le2}
\end{align}
Recall that as defined in Section~\ref{sec_f_life_table}, $\tilde{Z}_{L+s}$ is a time-shifted version of $\tilde{Z}_{L}$, where $\tilde{Z}_L^{(1)}$ (age) and $\tilde{Z}_L^{(2)}$ (calendar year) were shifted by $+s$. 
With ($\ref{s_le2}$), the equation (\ref{eq2}) leads to
\begin{align}
& \frac{\tilde{S}_{L \to O}(t|\tilde{Z}_L=\tilde{z}_L) - \alpha(\tilde{z}_L)\tilde{S}_{L \to O}(t|\tilde{Z}_L=\tilde{z}_L, \tilde{X}_L=1)}{\{1-\alpha(\tilde{z}_L)\}S_P(t|Z_D=\tilde{z}_L)} \nonumber \\
& = 1 - \int_{0}^{t} \left\{ 1 - 
\frac{S_{O}(t-s|Z_D=\tilde{z}_{L+s})}{S_{P}(t-s|Z_D=\tilde{z}_{L+s})}
\right\} d\tilde{F}_{L\to D}(s|\tilde{Z}_L=\tilde{z}_{L}, \tilde{X}_L=0).
\label{inteq1}
\end{align}
It is regarded as an integral equation with respect to $S_P(t| Z_D)$.

\subsection{Estimation of $S_P(t| Z_D)$ by solving the empirical integral equation}
\label{subsec_emp_SP}
In this subsection, we consider an empirical version of the integral equation ($\ref{inteq1}$), in which all the theoretical quantities are replaced with their empirical ones. 
We denote these empirical ones with the superscript of hat. 
For example, the empirical version of $\tilde{S}_{L \to O}(t|\tilde{Z}_L=\tilde{z}_L)$ is denoted by $\hat{S}_{L \to O}(t|\tilde{Z}_L=\tilde{z}_L)$. 
In the left hand side of ($\ref{inteq1}$), $\tilde{S}_{L \to O}(t|\tilde{Z}_L=\tilde{z}_L)$ is obtained from the life table. 
With the annual cancer incidence rate $IR(\tilde{z}_L)$, $\alpha(\tilde{Z}_L)$ is estimated by the method presented in Appendix B.1. 
Since $IR(\tilde{z}_L)$ is available only in an annual basis, we consider to estimate $S_P(t|Z_D)$ only at $t=0,1,2,\cdots$. 

As shown in Appendix A, $\tilde{S}_{L \to O}(t|\tilde{Z}_L=\tilde{z}_L, \tilde{X}_L=1)$ in the left hand side of ($\ref{inteq1}$) is represented as
\begin{align}
&\tilde{S}_{L \to O}(t|\tilde{Z}_L=\tilde{z}, \tilde{X}_L=1) \nonumber \\
&= 1 - \int_{0}^{\tilde{t}_L}{ \left\{ 1 - S_{O}(t+s|Z_D=\tilde{z}_{L-s}) \right\} d\tilde{F}_{D\to L}(s|\tilde{Z}_L=\tilde{z}, \tilde{X}_L=1) }, 
\label{s_lo}
\end{align}
under the assumptions (B-3) and (C-3). 
To handle the integral in the right hand side of (\ref{s_lo}), $S_{O}(t|Z_D=\tilde{z}_{L-s})$ for each $s = 0, 1, \cdots, \tilde{t}_L$ should be available over $(0, t+s)$. 
However, depending on the follow-up duration, it is not necessarily obtained from the cancer registry. 
To estimate over the interval, an extrapolation method with the Kaplan-Meier estimate is proposed in Appendix C. 
The method to estimate $\tilde{F}_{D\to L}(t|\tilde{Z}_L=\tilde{z}, \tilde{X}_L=1)$ is presented in Appendix B.2. 
Then, an estimator for $\tilde{S}_{L \to O}(t|\tilde{Z}_L=\tilde{z}, \tilde{X}_L=1)$ is given by 
\begin{align}
&\hat{S}_{L \to O}(t|\tilde{Z}_L=\tilde{z}, \tilde{X}_L=1) \nonumber \\
&= 1 - \int_{0}^{\tilde{t}_L}{ \left\{ 1 - \hat{S}_{O}(t+s|Z_D=\tilde{z}_{L-s}) \right\} d\hat{F}_{D\to L}(s|\tilde{Z}_L=\tilde{z}, \tilde{X}_L=1) }. \nonumber
\end{align}

Denote $\Delta \tilde{F}_k=\tilde{F}_{L\to D}(k|\tilde{Z}_L=\tilde{z}_{L}, \tilde{X}_L=0)-\tilde{F}_{L\to D}(k-1|\tilde{Z}_L=\tilde{z}_{L}, \tilde{X}_L=0)$ for $k=1,2, \cdots$ with $\tilde{F}_{L\to D}(0|\tilde{Z}_L=\tilde{z}_{L}, \tilde{X}_L=0)=0$. 
A method to estimate $\tilde{F}_{L\to D}(0|\tilde{Z}_L=\tilde{z}_{L}, \tilde{X}_L=0)$ is presented in Appendix B.3. 
Then, $\tilde{F}_{L\to D}(t|\tilde{Z}_L=\tilde{z}_{L}, \tilde{X}_L=0)=\sum_{k:k \le t} \Delta \tilde{F}_k$ and the integral equation (\ref{inteq1}) is represented by
\begin{align}
& \frac{\tilde{S}_{L \to O}(t|\tilde{Z}_L=\tilde{z}_L) - \alpha(\tilde{z}_L)\tilde{S}_{L \to O}(t|\tilde{Z}_L=\tilde{z}_L, \tilde{X}_L=1)}{\{1-\alpha(\tilde{z}_L)\}S_p(t|Z_D=\tilde{z}_L)} \nonumber \\
&= 1 - \sum_{k:k \le t}\left\{ 1 - 
\frac{S_{O}(t-k|Z_D=\tilde{z}_{L+k})}{S_{P}(t-k|Z_D=\tilde{z}_{L+k})}
\right\} \Delta \tilde{F}_k. 
\label{inteq2}
\end{align}
Set the right hand side of (\ref{inteq2}) as
\begin{align}
r(t|\tilde{z}_L) = 1 - \sum_{k:k \le t}{h_{\tilde{z}_L}(t,k) \Delta \tilde{F}_k}, \nonumber
\end{align}
where
\begin{align}
h_{\tilde{z}_L}(t,k)
&= 1 - \frac{S_{O}(t-k|Z_D=\tilde{z}_{L+k})}{S_{P}(t-k|Z_D=\tilde{z}_{L+k})}, \nonumber
\end{align}
and $h_{\tilde{z}_L}(k,k)=0$ for any $k$. 
Then (\ref{inteq2}) is represented as 
\begin{align}
r(t|\tilde{z}_L)
&=\frac{\tilde{S}_{L \to O}(t|\tilde{Z}_L=\tilde{z}_L) - \alpha(\tilde{z}_L)\tilde{S}_{L \to O}(t|\tilde{Z}_L=\tilde{z}_L, \tilde{X}_L=1)}{\{1-\alpha(\tilde{z}_L)\}S_P(t|Z_D=\tilde{z}_L)},
\label{inteq3}
\end{align}
Considering the equation (\ref{inteq2}) or (\ref{inteq3}) at $t=1, 2, \cdots, K$, one has the following system of the linear equations,
\begin{align}
&\left(
\begin{array}{c}
r(1|\tilde{z}_L) \\
r(2|\tilde{z}_L) \\
\vdots \\
r(K|\tilde{z}_L)
\end{array}
\right) 
=
\left(
\begin{array}{c}
1 \\
1 \\
\vdots \\
1
\end{array}
\right)
-
\left(
\begin{array}{ccccc}
0 & 0 & \cdots & 0 \\
h_{\tilde{z}_L}(2,1) & 0 & 0 & \vdots \\
\vdots & \vdots & \ddots & 0 \\
h_{\tilde{z}_L}(K,1) & h_{\tilde{z}_L}(K,2) & \cdots & 0
\end{array}
\right)
\left(
\begin{array}{c}
\Delta \tilde{F}_1\\
\Delta \tilde{F}_2 \\
\vdots, \\
\Delta \tilde{F}_K
\end{array}
\right) 
\label{inteq4}
\end{align}
The system of the linear equation ($\ref{inteq4}$) can be easily solved recursively replacing unknown theoretical quantities with their estimators as follows. 
The first equation of (\ref{inteq4}) is $r(1|\tilde{z}_L) =1$. 
Then, from the equation (\ref{inteq3}), $S_P(t|Z_D=\tilde{z}_L)$ at $t=1$ is estimated by 
\begin{align}
\hat{S}_P(1|Z_D=\tilde{z}_L) = \frac{\tilde{S}_{L \to O}(1|\tilde{Z}_L=\tilde{z}_L) - \hat{\alpha}(\tilde{z}_L)\hat{S}_{L \to O}(1|\tilde{Z}_L=\tilde{z}_L, \tilde{X}_L=1)}{\{1-\hat{\alpha}(\tilde{z}_L)\}}. \nonumber
\end{align}
The second equation of (\ref{inteq4}) is $r(2|\tilde{z}_L)=1 - h_{\tilde{z}_L}(2,1) \Delta \tilde{F}_1$. 
Set
\begin{align}
\hat{h}_{\tilde{z}_L}(2,1) = 1-\frac{\hat{S}_{O}(1|Z_D=\tilde{z}_{L+1})}{\hat{S}_{P}(1|Z_D=\tilde{z}_{L+1})} \ \ {\rm and} \ \
\hat{r}(2|\tilde{z}_L)=1 - \hat{h}_{\tilde{z}_L}(2,1) \Delta \hat{F}_1. \nonumber
\end{align}
Then, from (\ref{inteq3}), 
\begin{align}
& \hat{S}_P(2|Z_D=\tilde{z}_L) = \frac{\tilde{S}_{L \to O}(2|\tilde{Z}_L=\tilde{z}_L) - \hat{\alpha}(\tilde{z}_L)\hat{S}_{L \to O}(2|\tilde{Z}_L=\tilde{z}_L, \tilde{X}_L=1)}{\{1-\hat{\alpha}(\tilde{z}_L)\} \hat{r}(2|\tilde{z}_L)}. \nonumber
\end{align}
$S_P(t|Z_D=\tilde{z}_L)$ for $t \ge 3$ can be calculated recursively in a similar fashion by
\begin{align}
& \hat{S}_P(t|Z_D=\tilde{z}_L) = \frac{\tilde{S}_{L \to O}(t|\tilde{Z}_L=\tilde{z}_L) - \hat{\alpha}(\tilde{z}_L)\hat{S}_{L \to O}(t|\tilde{Z}_L=\tilde{z}_L, \tilde{X}_L=1)}{\{1-\hat{\alpha}(\tilde{z}_L)\} \hat{r}(t|\tilde{z}_L)}, \nonumber
\end{align}
where 
\begin{align}
\hat{r}(t|\tilde{z}_L) & =1 - \sum_{k: k\le t}{\hat{h}_{\tilde{z}_L}(t,k) \Delta \hat{F}_k} =1 - \sum_{k: k\le t}{\left\{ 1- \frac{\hat{S}_{O}(t-k|Z_D=\tilde{z}_{L+k})}{\hat{S}_{P}(t-k|Z_D=\tilde{z}_{L+k})} \right\} \Delta \hat{F}_k}. \nonumber
\end{align}
Following the above procedures, we estimate $S_P(t|Z_D=\tilde{z}_L)$ at $t=1,2,\cdots,K$, and the resulting estimator is denoted by $\hat{S}_P(t|Z_D=\tilde{z}_L)$. 
For $t$ other than $t=1,2,\cdots,K$, log-linear interpolation is applied. 
In Appendix D, a proof of consistency of $\hat{S}_P(t|Z_D=\tilde{z}_L)$ to $S_P(t|Z_D=\tilde{z}_L)$ is presented. 
Note that in the standard practice, $\hat{S}_{L\to O}(t|\tilde{Z}_L=\tilde{z}_L)$ is used as $S_P(t|Z_D=\tilde{z}_L)$ in (\ref{PP}). 
Instead, we propose to use $\hat{S}_P(t|Z_D=\tilde{z}_L)$ in calculating the $PP$ estimator.

\section{Simulation study}
\label{sec4}
We present the results of a simulation study to investigate the behavior of the proposed method. 
To generate the cancer registry data and life table, we consider the natural histories of subjects in a birth cohort as illustrated in Figure~\ref{f:relation}($c$). 
Each subject has the potential time to diagnosis of the cancer of interest after birth, $\tilde{t}_D$, and the potential time to death due to other causes after birth, denoted by $\tilde{t}_P$. 
If $\tilde{t}_D$ is shorter than $\tilde{t}_P$, the subject has the time to death due to cancer from the date of diagnosis, $\tilde{T}_{D\to E}$. 
The cancer registry data was constructed as the population registered at $\tilde{t}_D$, and it has the information of the baseline covariates at $\tilde{t}_D$ and time-to-death. 
The time-to-death due to any causes after diagnosis was calculated by $\tilde{T}_{D\to O}=\min(\tilde{T}_{D\to E}, \tilde{t}_P-\tilde{t}_D)$. 
From this information, the annual numbers of deaths from any cause, subjects diagnosed as cancer, and subjects in the population were calculated in each covariate. 
Then, the life tables and the annual cancer incidence rates were constructed.

We considered a cohort of 50,000 subjects born in 1960. 
We generated gender from the Bernoulli distribution with the probability of 0.5. 
$\tilde{t}_{D}$ and $\tilde{t}_{P}$ were generated under the four settings as follows; 
\begin{align}
{\rm Dataset} \ 1 & : \ 
\tilde{t}_{D} \sim Weibull(0.5\times 10^{-2}, \ 1), \ \tilde{t}_{P} \sim Weibull(1.0\times 10^{-2}, \ 2) \nonumber \\
{\rm Dataset} \ 2 & : \ 
\tilde{t}_{D} \sim Weibull(1.5\times 10^{-2}, \ 1), \ \tilde{t}_{P} \sim Weibull(1.0\times 10^{-2}, \ 2) \nonumber \\
{\rm Dataset} \ 3 & : \ 
\tilde{t}_{D} \sim LN(\log{65}, \ 2), \ \tilde{t}_{P} \sim LN(\log{75}, \ 2) \nonumber \\
{\rm Dataset} \ 4 & : \ 
\tilde{t}_{D} \sim LN(\log{65}, \ 1), \ \tilde{t}_{P} \sim LN(\log{75}, \ 2) \nonumber 
\end{align}
where $Weibull(\lambda, p)$ indicates the Weibull distribution with the hazard function of $\lambda p (\lambda t)^{p-1})$ and $LN(\mu, \sigma^2)$ indicates the Log-normal distribution. 
Datasets 1 and 3 had low cancer incidence and Datasets 2 and 4 had high cancer incidence. 
The covariates at the cancer diagnosis $\tilde{Z}_D=(age,year,gender)^{tr}$ were calculated by $(\tilde{t}_D, 1960 + \tilde{t}_D, gender)^{tr}$. 
$\tilde{T}_{D\to E}$ was generated from the exponential distribution with hazard rate $\lambda_E(t|\tilde{Z}_D)=\lambda \exp{\left\{ \beta^{tr} \tilde{Z}_D\right\}}$, where $\lambda=0.1\exp{ \left\{- \log{1.2} \times 60 / 7.5 - \log{0.95} \times (2000 - 1960)/15\right\} }$ and $\beta = ( \log{ 1.2 }/7.5, \log{ 0.95 }/15, \log{ 0.8 } )^{tr}$. 
The potential censoring time from diagnosis, $C$, was generated from the uniform distribution on $[0, 15]$. 
In this simulation, we focused on patients diagnosed from 60 to 74 years old, i.e. selected by $\tilde{t}_D \in [60, 75)$. 
We simulated 1,000 datasets in each setting. 
The true net survival function $S_E(t)=E_{Z_D}\left[\exp( -t\lambda_E(t|Z_D) )\right]$ was calculated by the average of $\exp( -t\lambda_E(t|Z_D) )$ over $n=500,000$. 

Table~\ref{t:sim_t2} displays the number of cancer patients and events in each of four datasets. 
To apply the proposed method, we calculated the Kaplan-Meier estimators for each subpopulation with $Z_D$ to estimate $S_O(t|Z_D)$. 
For the log-linear extrapolation of $S_O(t|Z_D)$ beyond the end of follow-up in (\ref{s_lo}), we used $H=4$ and $H=10$. 
Table~\ref{t:sim1} showed the empirical biases and root mean squared errors (rMSEs) of estimates for 3-, 5-, 7-, and 10-years net survivals. 
In all datasets, the $PP$ estimator had considerable biases particularly in the case of high incidence (Datasets 3 and 4). 
The proposed method had negligible biases for all time points and the rMSEs of the proposed methods were smaller than that of the $PP$ estimator. 
No substantial differences in estimation accuracy were observed between extrapolation using the $H=4$ and with $10$.

\section{Illustration}
\label{sec_example}
We illustrated our proposed method by analyzing two cancer sites from the National Cancer Registry at the Office for National Statistics. We focused on a subgroup of all adult aged 65–74 years, who was diagnosed as colon or prostate cancers from 1990 to 2000 in London, England. All patients were followed up to 15 years after diagnosis. 
For colon cancer, 55,033 patients were included and 48,549 among them died until the end of follow-up. For prostate cancer, 71,419 patients were included, and 62,422 patients died. 
The data were analyzed by cancer sites (colon and prostate). 
To apply the $PP$ estimator, set $Z_D=(age, year, gender)^{tr}$ in colon cancer and $Z_D=(age, year)^{tr}$ in prostate cancer. 

To calculate $\tilde{S}_{L\to O}(t|\tilde{Z}_L)$, we used the population life-table of England, which gives annual mortality from 1981 to 2015 by age and gender.
To calculate the incidence rate, we used the information of the number of populations by age, gender, and calendar year, which are available from England Population Estimates 1971 to 2014 (\url{https://www.ons.gov.uk/peoplepopulationandcommunity}). 
The survival function $S_O(t|Z_D)$ was estimated by the Kaplan-Meier method applied to the subpopulation by $Z_D$. 
Extrapolation of $S_O(t|Z_D)$ required in (\ref{s_lo}) was made by the method in Appendix C with $H=4$ and $10$. 

Figure~\ref{f:england1}(A) shows the net survival estimated by the $PP$ estimator with the standard approach of the population life-table and that with the proposed adjustment ($H=4$ or $10$) for colon cancer. 
Correspondingly, the estimated net survival probabilities at selected time points are shown in Table~\ref{t:england}. 
Recall that in the standard approach, $\tilde{S}_{L\to O}(t|\tilde{Z}_L)$ is used as $S_P(t|Z_D)$, whereas in the proposed adjustment $S_P(t|Z_D)$ is estimated. 
To see the magnitude of the difference between both approaches, in Figure~\ref{f:england1}(B), $\tilde{S}_{L\to O}(t|\tilde{Z}_L=(65,1990,male))$ and $\hat{S}_P(t|Z_D=(65,1990,male))$ are plotted, which are used for estimates of $S_P(t|Z_D=(65, 1990,male))$ in (\ref{PP}) in the standard method and the proposed one, respectively. 
Figure~\ref{f:england1}(C) shows $\tilde{S}_{L \to E}(t|\tilde{Z}_L=(65, 1990,male), \tilde{X}_L=0)$, and Figure~\ref{f:england1}(D) shows $\alpha(\tilde{Z}_L)$ by age. 
Recall that $\tilde{S}_{L \to E}(t|\tilde{Z}_L, \tilde{X}_L=0)=1$ for any $t$ and $\alpha(\tilde{Z}_L)=0$ hold if the life table did not include cancer patients or cancer deaths. 
Figure~\ref{f:england1}(C) indicated only small inclusion of cancer death in the life tables, and Figure~\ref{f:england1}(D) showed that the inclusion of cancer patients is also minor. 
Correspondingly, as seen in Figure~\ref{f:england1}(B), estimated $S_P(t|Z_D)$ were slightly different with the proposed adjustment. 
Then, as seen in Figure~\ref{f:england1}(A), the $PP$ estimates of net survival were modified by the proposed method by 0.4 to 0.6\% at any time points. 

Results for prostate cancer are presented in Figure~\ref{f:england2}. 
On the contrary to colon cancer, there was a rather higher impact by the inclusion of prostate cancer deaths in the life table as seen in Figure~\ref{f:england2}(A); the $PP$ estimates with the proposed method were 0.8 to 1.6\% lower than those with the standard use of the life table. 
As seen in Figure~\ref{f:england2}(B), the proposed method made a certain amount of adjustment in estimation of $S_P(t|Z_D=(65, 1990))$. 
As shown in Figure~\ref{f:england2}(C) and \ref{f:england2}(D), the inclusion of cancer population and cancer deaths in the life table would not be ignorable.

\section{Discussion}
\label{s:discuss}
When analyzing population-based cancer registry data, external life tables (from the general population) are commonly utilized to estimate cancer-related survival measures (such as net survival) within the relative survival data setting. 
Such estimation assumes the absence of cancer patients and cancer deaths in the life tables, which cannot be fully met. 
Although the issue is generally ignored by assuming a minor impact on the estimation of net survival, a sensitivity analysis method to address it was also proposed by \citet{Talback2011}. 
Their sensitivity analysis requires information on the number of cancer deaths in the general population, an information which is not available in the data usually collected by cancer registries. 
In this paper, we demonstrate how to address this problem with a method based on an easily tractable integral equation.
In contrast to the approach introduced in \citet{Talback2011}, our method requires only information contained in the standard cancer registry datasets. 

Our method is easily extendable to various measures other than net survival, such as relative survival ratio or crude probability of death. 
The relative survival is defined as $S_R(t)=S_O(t)/S_P(t)$, which is the ratio between overall survival in the cancer patients and that for the general population. 
The Ederer I ($E1$) estimator \cite{Ederer1961,Perme2012} of $\Lambda_R(t)=- \log{S_R(t)}$, which is the consistent estimator of the relative survival ratio, is defined by
\begin{eqnarray}
\hat{\Lambda}_R^{E1} (t)
= \int _{0}^{t} \frac{\sum _{i=1}^{n} dN_i (u)}{\sum _{j=1}^{n} Y_j (u) } - \int _{0}^{t} \frac{\sum _{i=1}^{n} \tilde{S}_{L\to O}(u|Z_{D,i}) d\tilde{\Lambda}_{L\to O} (u|Z_{D,i})}{\sum _{j=1}^{n} \tilde{S}_{L\to O}(u|Z_{D,j}) }. 
\label{E1}
\end{eqnarray} 
The $E1$ estimator is a consistent estimator of $\Lambda_R(t)$ under the condition $C\perp \{T_O, Z_D\}$ (independent censoring). 
With the $E1$ estimator, one can replace $\tilde{S}_{L\to O}(t|Z_{D})$ and $\tilde{\Lambda}_{L\to O} (t|Z_{D})$ in the equation (\ref{E1}) with the adjusted version. 
The crude probability of death is defined by $F_{CPD}(t)=\int_0^t{ S_O(u)\lambda_E^*(u)du }$, where $\lambda_E^*(u)=\lim_{h \to 0}\Pr(t<T_E \le t+h|T_O \ge t)/h$ is a cause specific hazard due to cancer. 
An estimator of the crude probability of death is defined by $\hat{F}_{CPD}(t)=\int_0^t{ \hat{S}_O(u)d\hat{\Lambda}_E^*(u) }$,\cite{Cronin2000, Perme2012, Perme2018} where $\hat{S}_O(t)$ is the estimator of the overall survival, i.e., its cumulative hazard function is estimated as the Nelson-Aalen estimator, and 
\begin{eqnarray}
\hat{\Lambda}_E^{*} (t)
= \int _{0}^{t} \frac{\sum _{i=1}^{n} dN_i (u)}{\sum _{j=1}^{n} Y_j (u) } - \int _{0}^{t} \frac{\sum _{i=1}^{n} Y_i (u) d\Lambda_P (u|Z_{D,i})}{\sum _{j=1}^{n} Y_j (u) }. 
\label{E2}
\end{eqnarray} 
This estimator is a consistent under the independent censoring assumption $C \perp \{T_O, T_P, Z_D\}$. 
For this estimator, $\Lambda_P (u|Z_{D})$ needs to be replaced with the adjusted version. 

\citet{Ederer1961}, \citet{Esteve1994}, and others have stated that the presence of cancer death in the life table had a minimal impact on the estimation of these cancer survival measures. 
Although our illustration supported this in particular in the case of low cancer incidence, it is not necessarily true if the incidence rate is not low and the incidence rate of some cancer types may increase in the future.\cite{Siegel2016} 
Then, it is valuable to have tools to address the issue quantitatively. 
Furthermore, the net survival and other related survival measures have attracted attention in the context of human immunodeficiency virus (HIV) cohorts \cite{Marston2005, Marston2007, Bhaskaran2008, Marston2011} or of cardiovascular diseases \cite{Nelson2008, Lantelme2022}. 
Because of the dramatic improvement of prognosis among individuals infected with HIV following the widespread introduction of highly active antiretroviral therapy (HAART), it became important to account for the competing risks of death from other causes when estimating survival from HIV. 
In the absence of accurate information on the cause of death, methods developed for the relative survival setting can be applied. 
Because HIV prevalence is very high in some African regions (for example exceeding 10\% in the 15-49 age group in South Africa and Botswana), it is crucial to address the presence of HIV patients and HIV deaths in the life tables. 
Similarly, the high prevalence and mortality of cardiovascular diseases in many populations are likely to violate the assumptions underlying survival estimation approaches within the relative survival setting. 
In such situations, the proposed method would play very important roles.

\section*{Acknowledgment}
The first author's research was partly supported by Grant-in-Aid for Early-Career Scientists (20K19754) from the Ministry of Education, Science, Sports and Technology of Japan.
The second author's research was partly supported by Grant-in-Aid for Challenging Exploratory Research (16K12403) and for Scientific Research (16H06299, 18H03208) from the Ministry of Education, Science, Sports and Technology of Japan. 
The third author's research was partly supported by Cancer Research UK (Reference C7923/A18525). 
Computational calculations were performed at the Institute of Medical Science (the University of Tokyo). 

{\it Conflict of Interest}: None declared.

\clearpage

\clearpage


\begin{figure}
\centering\includegraphics[width=150mm,height=150mm, angle=-90]{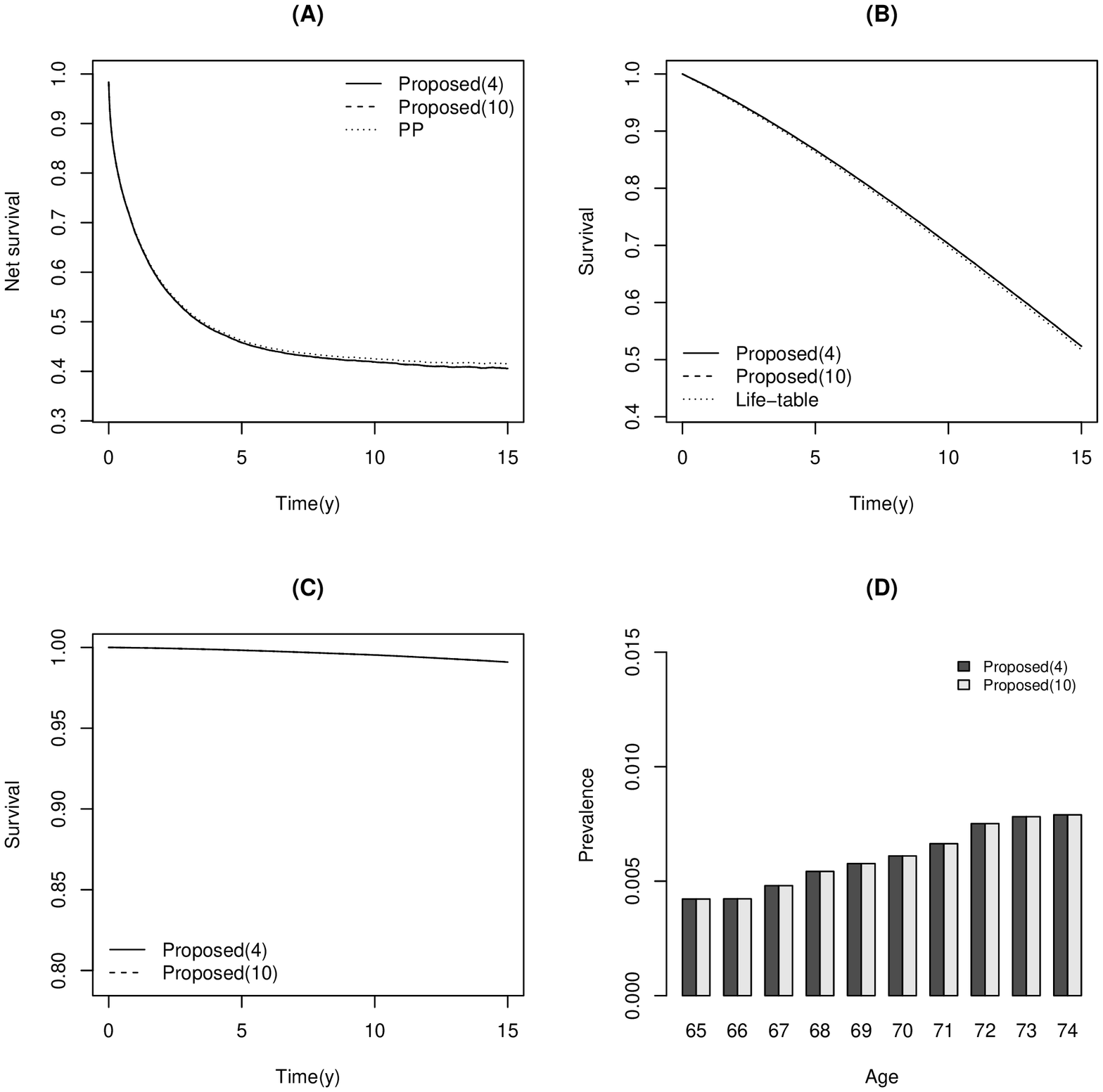}
\caption{Plots of estimated survival curves and prevalence rates for colon cancer patients diagnosed at age 64-74 in England; (A) shows the net survival curves by the proposed methods with $H=4$ and $10$ and the Pohar-Perme estimator; (B) plot the population survival curve from the life table and the non-cancer survival estimated by the proposed methods; (C) plots the cancer survival function for the non-cancer patient at the registration to the life table; (D) shows the estimated prevalence rates of male population at each age in 1990. (B) and (C) show the survival functions for male population of 65 years old in 1990. }
\label{f:england1}
\end{figure}


\clearpage


\begin{figure}
\centering\includegraphics[width=150mm,height=150mm, angle=-90]{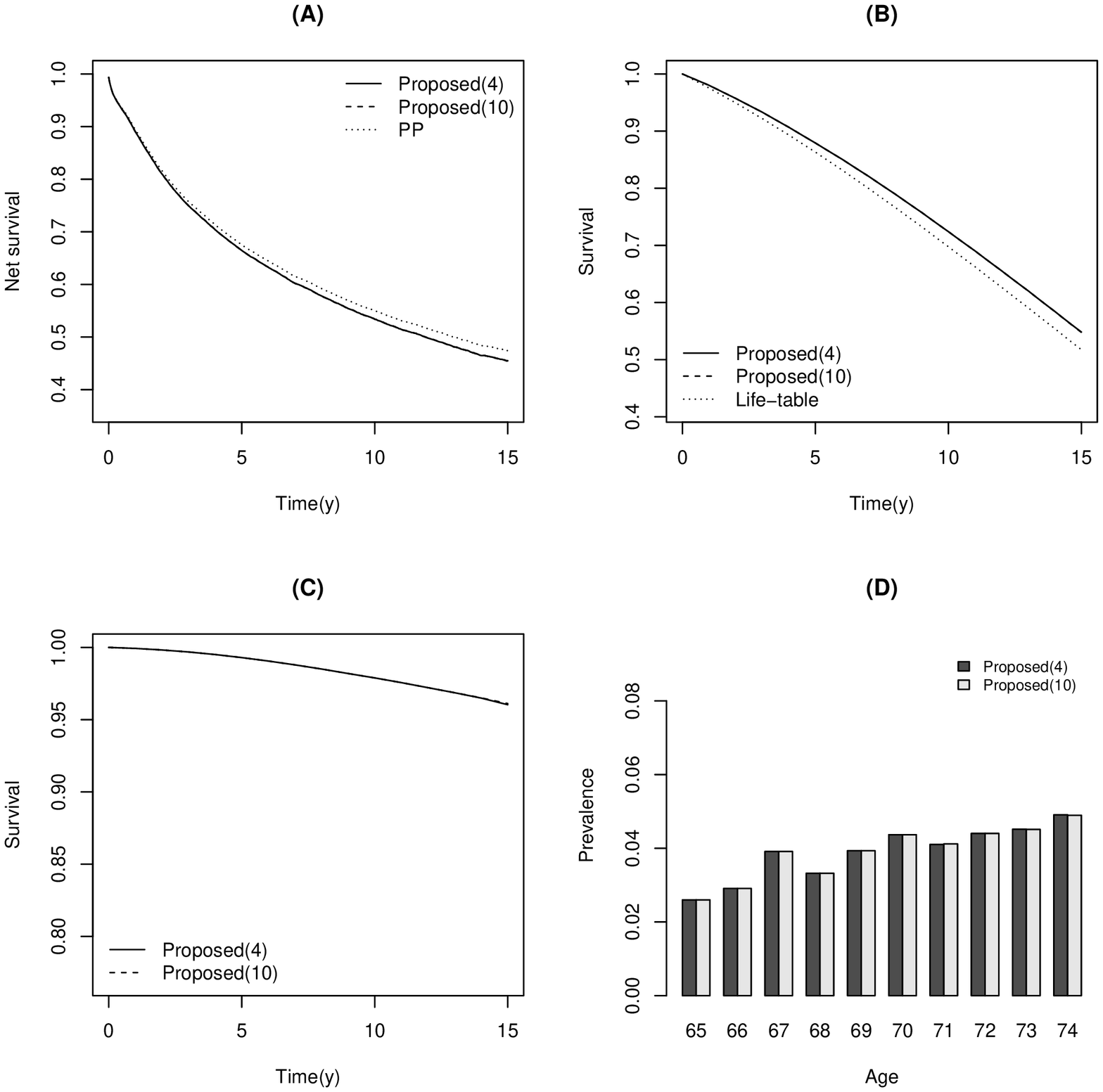}
\caption{Plots of estimated survival curves and prevalence rates for prostate cancer patients diagnosed at age 65-74 in England; (A) shows the net survival curves by the proposed methods with $H=4$ and $10$ and the Pohar-Perme estimator; (B) plot the population survival curve from the life table and the non-cancer survival estimated by the proposed methods; (C) plots the cancer survival function for the non-cancer patient at the registration to the life table; (D) shows the estimated prevalence rates at each age in 1990. (B) and (C) show the survival functions for 65 years old in 1990.}
\label{f:england2}
\end{figure}


\clearpage

\begin{table}
\centering
\caption{
Summary of simulated cancer registry data: Medians with ranges of the number of cancer patients diagnosed from 60 to 74 years old and of the number of events across the 1,000 simulation data are presented. The dataset 1 and 2 are by the low and high incidence Weibull models, respectively. The dataset 3 and 4 are by the low and high incidence log-normal model, respectively. 
}
\label{t:sim_t2}
\begin{tabular}{ccccccc}
\hline
Dataset & Patients(Median[Range]) & Events(Median[Range]) \\
\hline
1 & 1699 [1572, 1836] & 870 [771, 970] \\
2 & 2605 [2431, 2753] & 1329.5 [1196, 1426] \\
3 & 1160 [1058, 1272] & 565 [501, 634] \\
4 & 2320 [2162, 2449] & 1127 [1026, 1238] \\
\hline
\end{tabular}
\end{table}

\clearpage

\begin{sidewaystable}
\centering
\caption{
Results of the simulation studies for the evaluation of the proposed estimators with the use of 4-points and 10-points near the end of follow-up: Average of estimates, percentage bias(\%), and rMSE (empirical root of mean squared error) across the 1,000 simulation data are presented in each year. The dataset 1and 2 are by the low and high incidence Weibull models, respectively. The dataset 3 and 4 are by the low and high incidence log-normal model, respectively. 
}
\label{t:sim1}
\begin{tabular}{cccccccccccc}
\hline
& & & \multicolumn{3}{c}{Proposed($H=4$)} & \multicolumn{3}{c}{Proposed($H=10$)} & \multicolumn{3}{c}{Pohar-Perme} \\
Dataset & Year & True & Ave & \%Bias & rMSE($\times 100$) & Ave & \%Bias & rMSE($\times 100$) & Ave & \%Bias & rMSE($\times 100$) \\
\hline
1&3&0.748&0.748&-0.05&1.23&0.747&-0.17&1.24&0.761&1.76&1.77\\
&5&0.617&0.617&0.02&1.48&0.616&-0.17&1.48&0.636&3.02&2.37\\
&7&0.51&0.51&0.06&1.69&0.509&-0.18&1.69&0.531&4.23&2.74\\
&10&0.383&0.384&0.22&1.87&0.383&-0.06&1.87&0.406&6.06&3.01\\
\hline
2&3&0.749&0.751&0.31&1.16&0.749&-0.01&1.14&0.789&5.28&4.08\\
&5&0.618&0.622&0.62&1.4&0.619&0.14&1.35&0.674&9.05&5.73\\
&7&0.511&0.517&1.15&1.55&0.514&0.52&1.47&0.578&13.1&6.84\\
&10&0.384&0.392&2.08&1.78&0.389&1.3&1.67&0.458&19.21&7.57\\
\hline
3&3&0.748&0.756&0.95&1.65&0.755&0.85&1.62&0.77&2.85&2.57\\
&5&0.617&0.628&1.69&2.01&0.627&1.53&1.97&0.648&4.89&3.46\\
&7&0.51&0.522&2.37&2.23&0.521&2.18&2.18&0.545&6.90&3.99\\
&10&0.383&0.396&3.34&2.42&0.395&3.14&2.38&0.421&9.88&4.34\\
\hline
4&3&0.749&0.752&0.36&1.15&0.750&0.17&1.13&0.781&4.29&3.37\\
&5&0.618&0.623&0.77&1.37&0.621&0.48&1.32&0.664&7.41&4.74\\
&7&0.51&0.516&1.1&1.51&0.514&0.71&1.45&0.564&10.50&5.54\\
&10&0.384&0.391&2.03&1.76&0.390&1.57&1.69&0.443&15.49&6.17\\
\hline
\end{tabular}
\end{sidewaystable}


\clearpage

\begin{table}
\centering
\caption{
Estimates of the 3-year, 5-year, 7-year, and 10-year net survivals for patients with colon and prostate cancers (the number of events: 55,033 in colon cancer and 71,419 in prostate, the number of patients: 48,549 in colon cancer and 62,422 in prostate) in cancer registry data in England. 
}
\label{t:england}
\begin{tabular}{ccccccc}
\hline
& & \multicolumn{2}{c}{Proposed} & Pohar-Perme \\
Site & Year & $H=4$ & $H=10$ & \\
\hline
colon & 3 & 0.517 & 0.517 & 0.521 \\
& 5 & 0.458 & 0.458 & 0.462 \\
& 7 & 0.433 & 0.433 & 0.438 \\
& 10 & 0.419 & 0.419 & 0.425 \\
\hline
prostate & 3 & 0.750 & 0.750 & 0.758 \\
& 5 & 0.665 & 0.665 & 0.675 \\
& 7 & 0.602 & 0.602 & 0.615 \\
& 10 & 0.534 & 0.534 & 0.550 \\
\hline
\end{tabular}
\end{table}

\clearpage

\appendix

\section*{Appendix}
\section*{A: Derivations of (5) and (7)}
\label{Ap2}

Recall that $\tilde{T}_{L \to E}=\tilde{T}_{L \to D}+\tilde{T}_{D \to E}$ for a non-cancer subject at $\tilde{t}_L$ ($\tilde{X}_L=0$). Under the assumption (B-2), by using the convolution formula (Chapter 2 in Durrett, 2010) for a sum of independent random variable, it holds that
\begin{align}
&\tilde{S}_{L \to E}(t|\tilde{Z}_L=\tilde{z}_L, \tilde{X}_L=0) = \Pr(\tilde{T}_{L \to D}+\tilde{T}_{D \to E}>t|\tilde{Z}_L=\tilde{z}_L, \tilde{X}_L=0)\nonumber \\
&=1 - \Pr(\tilde{T}_{L \to D}+\tilde{T}_{D \to E}\le t|\tilde{Z}_L=\tilde{z}_L, \tilde{X}_L=0) \nonumber \\
&=1 - \int_{0}^{t}{\left\{ 1 - \tilde{S}_{D\to E}(t-s|\tilde{Z}_D=\tilde{z}_{L+s}, \tilde{X}_L=0) \right\} d\tilde{F}_{L\to D}(s|\tilde{Z}_L=\tilde{z}_L, \tilde{X}_L=0)}.
\nonumber
\end{align} 
Note that the origin of $\tilde{S}_{D\to E}(t-s|\tilde{Z}_D=\tilde{z}_{L+s}, \tilde{X}_L=0)$ in the integrand is the date when a subject is diagnosed as cancer. Thus, the covariate $\tilde{Z}_D$ was shifted by $s$ from $\tilde{Z}_L$. From (C-1), $S_E(t|Z_D=\tilde{z}_{L+s})=\tilde{S}_{D\to E}(t|\tilde{Z}_D=\tilde{z}_{L+s}, \tilde{X}_L=0)$. Under (A-1), $S_{E}(t-s|Z_D=\tilde{z}_{L+s}) = S_{O}(t-s|Z_D=\tilde{z}_{L+s})/S_{P}(t-s|Z_D=\tilde{z}_{L+s})$ holds. 
Then, we have (5).

For the subjects with $\tilde{X}_L=1$, $\tilde{T}_{L \to O}=\tilde{T}_{D \to O}-\tilde{T}_{D \to L}$. Then, the convolution technique (Chapter 2 in Durrett, 2010) applied under (B-3) gives 
\begin{align}
&\tilde{S}_{L \to O}(t|\tilde{Z}_L=\tilde{z}_L, \tilde{X}_L=1) \nonumber \\
&= 1 - \int_{0}^{\tilde{t}_L}{ \left\{ 1 - \tilde{S}_{D \to O}(t+s|\tilde{Z}_L=\tilde{z}_L, \tilde{X}_L=1) \right\} d\tilde{F}_{D\to L}(s|\tilde{Z}_L=\tilde{z}_L, \tilde{X}_L=1) } \nonumber \\
&= 1 - \int_{0}^{\tilde{t}_L}{ \left\{ 1 - \tilde{S}_{D \to O}(t+s|\tilde{Z}_D=\tilde{z}_{L-s}, \tilde{X}_L=1) \right\} d\tilde{F}_{D\to L}(s|\tilde{Z}_L=\tilde{z}_L, \tilde{X}_L=1) } \nonumber \\
&= 1 - \int_{0}^{\tilde{t}_L}{ \left\{ 1 - S_{O}(t+s|Z_D=\tilde{z}_{L-s}) \right\} d\tilde{F}_{D\to L}(s|\tilde{Z}_L=\tilde{z}_L, \tilde{X}_L=1) }. \nonumber
\end{align}
where the last equality holds from (C-3). 
It completes the derivation of (7). 

\section*{B: Estimation of the prevalence rate $\alpha(\tilde{z}_L)$ and the related quantities}
\label{use_of_ir}
In this appendix, we explain how to estimate $\alpha(\tilde{z}_L)$, $\tilde{F}_{D \to L}(t|\tilde{Z}_L=\tilde{z}_L,\tilde{X}_L=1)$ in (7), and $\tilde{F}_{L \to D}(t|\tilde{Z}_L=\tilde{z}_L,\tilde{X}_L=0)$ in (6) with the information on the annual incidence rates $IR(\tilde{z}_L)$ and the cancer registry data. 
They are summarized as in the subsections B.1, B.2, and B.3, respectively. 
Denote the joint density of $\tilde{t}_D$ and $\tilde{T}_{D\to O}$ given the covariates $\tilde{Z}_L=\tilde{z}_L$ by $\tilde{f}_{\tilde{t}_D,\tilde{T}_{D\to O}}(u,s|\tilde{Z}_L=\tilde{z}_L)$. 
Let $\tilde{f}_{\tilde{T}_{D\to O}|\tilde{t}_D}(s|\tilde{t}_D=u, \tilde{Z}_L=\tilde{z}_L)$ be the conditional density function of $\tilde{T}_{D\to O}$ given $\tilde{t}_D=u$ and $\tilde{Z}_L=\tilde{z}_L$, and $\tilde{f}_{\tilde{t}_D}(u|\tilde{Z}_L=\tilde{z}_L)$ be that of $\tilde{t}_D$ given $\tilde{Z}_L=\tilde{z}_L$. 

\subsection*{B.1. The prevalence rate $\alpha(\tilde{z}_L)$}
\label{use_of_ir_for_alpha}
The prevalence rate $\alpha(\tilde{z}_L)$ is represented by 
\begin{align}
\alpha(\tilde{z}_L) 
&= \Pr(\tilde{X}_{L} = 1|\tilde{Z}_L=\tilde{z}_{L}) 
= \int_{0}^{\tilde{t}_L}{ \int_{\tilde{t}_L-u}^{\infty}{ \tilde{f}_{\tilde{t}_D,\tilde{T}_{D\to O}}(u,s|\tilde{Z}_L=\tilde{z}_L) ds } du } \nonumber \\
&= \sum_{u'=0}^{\tilde{t}_L-1}{ \int_{u'}^{u'+1}{ \int_{\tilde{t}_L-u}^{\infty}{ \tilde{f}_{\tilde{T}_{D\to O}|\tilde{t}_D}(s|\tilde{t}_D=u, \tilde{Z}_L=\tilde{z}_L) ds } \tilde{f}_{\tilde{t}_D}(u|\tilde{Z}_L=\tilde{z}_L) du } }. \nonumber
\end{align}
Suppose $\tilde{t}_D \in [u',u'+1)$. 
Then, from the assumption that the covariates remains constant in each year, $\tilde{Z}_{D}=\tilde{Z}_{L-\tilde{t}_L+u'}$ (see Figure 1(c)). 
$\{ \tilde{Z}_L=\tilde{z}_L \}$ is the same event as $\{ \tilde{Z}_{L-\tilde{t}_L+u'}=\tilde{z}_{L-\tilde{t}_L+u'} \}$, and $\{ \tilde{t}_D=u, \tilde{Z}_{L-\tilde{t}_L+u'}=\tilde{z}_{L-\tilde{t}_L+u'} \}$ and $\{ \tilde{t}_D=u, \tilde{Z}_{D}=\tilde{z}_{L-\tilde{t}_L+u'}\}$ are also the same event. 
Since $\tilde{Z}_{D}$ contains age at diagnosis, $\tilde{t}_D$ is constant conditional on $\tilde{Z}_{D}$. 
Therefore, we have
\begin{align}
\alpha(\tilde{z}_L)
&= \sum_{u'=0}^{\tilde{t}_L-1}{ \int_{u'}^{u'+1}{ \int_{\tilde{t}_L-u}^{\infty}{ \tilde{f}_{\tilde{T}_{D\to O}|\tilde{t}_D}(s|\tilde{t}_D=u, \tilde{Z}_{L-\tilde{t}_L+u'}=\tilde{z}_{L-\tilde{t}_L+u'}) ds } } } \nonumber \\
& \qquad \qquad \qquad \qquad \qquad \qquad \qquad \times \tilde{f}_{\tilde{t}_D}(u|\tilde{Z}_{L-\tilde{t}_L+u'}=\tilde{z}_{L-\tilde{t}_L+u'}) du \nonumber \\
&= \sum_{u'=0}^{\tilde{t}_L-1}{ \int_{u'}^{u'+1}{ \int_{\tilde{t}_L-u}^{\infty}{ \tilde{f}_{\tilde{T}_{D\to O}}(s|\tilde{Z}_{D}=\tilde{z}_{L-\tilde{t}_L+u'}) ds } \tilde{f}_{\tilde{t}_D}(u|\tilde{Z}_{L-\tilde{t}_L+u'}=\tilde{z}_{L-\tilde{t}_L+u'}) du } } \nonumber \\
&= \sum_{u'=0}^{\tilde{t}_L-1}{ \int_{u'}^{u'+1}{ \tilde{S}_{D \to O}(\tilde{t}_L-u|\tilde{Z}_D=\tilde{z}_{L-\tilde{t}_L+u'}) \tilde{f}_{\tilde{t}_D}(u|\tilde{Z}_{L-\tilde{t}_L+u'}=\tilde{z}_{L-\tilde{t}_L+u'}) du } }, 
\label{alpha_int}
\end{align}
where the last equality holds because $\int_{\tilde{t}_L-u}^{\infty}{ \tilde{f}_{\tilde{T}_{D\to O}}(s|\tilde{Z}_{D}=\tilde{z}_{L-\tilde{t}_L+u'}) ds }$ is the conditional probability of $\{ \tilde{T}_{D\to O} > \tilde{t}_L-u \}$ given $\tilde{Z}_{D}=\tilde{z}_{L-\tilde{t}_L+u'}$. 
On $u \in [u',u'+1)$, we approximate $\tilde{S}_{D \to O}(\tilde{t}_L-u|\tilde{Z}_D=\tilde{z}_{L-\tilde{t}_L+u'}) \approx \tilde{S}_{D \to O}(\tilde{t}_L-u'|\tilde{Z}_D=\tilde{z}_{L-\tilde{t}_L+u'})$. 
From the assumption (C-3), $\tilde{S}_{D \to O}(\tilde{t}_L-u'|\tilde{Z}_D=\tilde{z}_{L-\tilde{t}_L+u'})=S_O(\tilde{t}_L-u'|Z_D=\tilde{z}_{L-\tilde{t}_L+u'})$. 
Then, summand of (\ref{alpha_int}) is
\begin{align}
&\int_{u'}^{u'+1}{ \tilde{S}_{D \to O}(\tilde{t}_L-u|\tilde{Z}_D=\tilde{z}_{L-\tilde{t}_L+u'}) \tilde{f}_{\tilde{t}_D}(u|\tilde{Z}_{L-\tilde{t}_L+u'}=\tilde{z}_{L-\tilde{t}_L+u'}) du } \nonumber \\
&\approx S_O(\tilde{t}_L-u'|Z_D=\tilde{z}_{L-\tilde{t}_L+u'})\int_{u'}^{u'+1}{ \tilde{f}_{\tilde{t}_D}(u|\tilde{Z}_{L-\tilde{t}_L+u'}=\tilde{z}_{L-\tilde{t}_L+u'}) du }. \nonumber 
\end{align}
It holds that
\begin{align}
&\int_{u'}^{u'+1}{ \tilde{f}_{\tilde{t}_D}(u|\tilde{Z}_{L-\tilde{t}_L+u'}=\tilde{z}_{L-\tilde{t}_L+u'}) du } \nonumber \\
&= \int_{u'}^{u'+1}{ \tilde{f}_{\tilde{t}_D}(u|\tilde{Z}_{L-\tilde{t}_L+u'}=\tilde{z}_{L-\tilde{t}_L+u'}, \tilde{X}_{L-\tilde{t}_L+u'}=0) du} \nonumber \\
&\qquad \times \Pr(\tilde{X}_{L-\tilde{t}_L+u'}=0|\tilde{Z}_{L-\tilde{t}_L+u'}=\tilde{z}_{L-\tilde{t}_L+u'}) \nonumber \\
&= IR(\tilde{z}_{L-\tilde{t}_L+u'}) \left\{ 1 - \alpha(\tilde{z}_{L-\tilde{t}_L+u'})\right\}. 
\label{int_up_up1}
\end{align}
Then, (\ref{alpha_int}) is represented as
\begin{align}
\alpha(\tilde{z}_L)
&\approx \sum_{u'=0}^{\tilde{t}_L-1}{ S_O(\tilde{t}_L-u'|Z_D=\tilde{z}_{L-\tilde{t}_L+u'}) IR(\tilde{z}_{L-\tilde{t}_L+u'}) \left\{ 1 - \alpha(\tilde{z}_{L-\tilde{t}_L+u'})\right\} }. \nonumber
\end{align}
By changing variables, $\alpha(\tilde{z}_L)$ can be approximated by 
\begin{align}
\alpha(\tilde{z}_L) \approx \sum_{s=1}^{\tilde{t}_L}{ S_O(s|Z_D=\tilde{z}_{L-s}) IR(\tilde{z}_{L-s}) \left\{ 1 - \alpha(\tilde{z}_{L-s})\right\} }, \nonumber 
\end{align} 
and we can estimate it recursively as follows. 
Denote the estimator of $\alpha(\tilde{z}_L)$ by $\hat{\alpha}(\tilde{z}_L)$. 
Recall that $\tilde{z}_L$ is the real value of $\tilde{Z}_L=(\tilde{Z}_L^{(1)}, \tilde{Z}_L^{(2)}, \tilde{Z}_L^{(others)})$ (see Section 3 and Figure 1). 
When $\tilde{t}_L=0$, suppose that $\alpha(\tilde{z}_L=(0,\tilde{z}_L^{(2)},\tilde{z}_L^{(others)}))=0$. 
Set $\tilde{t}_L=1$, that is, the age at diagnosis is 1 years old, $\tilde{z}_L^{(1)}=1$. 
Then, the estimator of $\alpha(\tilde{z}_L=(1,\tilde{z}_L^{(2)},\tilde{z}_L^{(others)}))$ is given by
\begin{align}
\hat{\alpha}(\tilde{z}_L=(1,\tilde{z}_L^{(2)},\tilde{z}_L^{(others)})) = \hat{S}_O(1|Z_D=\tilde{z}_{L-1}) IR(\tilde{z}_{L-1}) \left\{ 1 - \hat{\alpha}(\tilde{z}_{L-1})\right\}, \nonumber 
\end{align} 
with $\tilde{z}_{L-1}=(0,\tilde{z}_L^{(2)}-1,\tilde{z}_L^{(others)})$. 
Set $\tilde{t}_L=2$. 
By using the estimators $\hat{\alpha}(\tilde{z}_{L-1}=(1,\tilde{z}_L^{(2)}-1,\tilde{z}_L^{(others)}))$ and $\hat{\alpha}(\tilde{z}_{L-2}=(0,\tilde{z}_L^{(2)}-2,\tilde{z}_L^{(others)}))$, $\alpha(\tilde{z}_L=(2,\tilde{z}_L^{(2)},\tilde{z}_L^{(others)}))$ is estimated by
\begin{align}
\hat{\alpha}(\tilde{z}_L=(2,\tilde{z}_L^{(2)},\tilde{z}_L^{(others)})) 
&= \hat{S}_O(1|Z_D=\tilde{z}_{L-1}) IR(\tilde{z}_{L-1}) \left\{ 1 - \hat{\alpha}(\tilde{z}_{L-1})\right\} \nonumber \\
&+ \hat{S}_O(2|Z_D=\tilde{z}_{L-2}) IR(\tilde{z}_{L-2}) \left\{ 1 - \hat{\alpha}(\tilde{z}_{L-2})\right\}. \nonumber
\end{align} 
For $\tilde{t}_L \ge 3$, $\alpha(\tilde{z}_L)$ is estimated recursively by 
\begin{align}
\hat{\alpha}(\tilde{z}_L) \approx \sum_{s=1}^{\tilde{t}_L}{ \hat{S}_O(s|Z_D=\tilde{z}_{L-s}) IR(\tilde{z}_{L-s}) \left\{ 1 - \hat{\alpha}(\tilde{z}_{L-s})\right\} }, \nonumber 
\end{align} 

\subsection*{B.2. $\tilde{F}_{D \to L}(t|\tilde{Z}_L=\tilde{z}_L,\tilde{X}_L=1)$ in (7)}
\label{use_of_ir_for_F_DL}
Recall that $\alpha(\tilde{z}_L)=\Pr(\tilde{X}_L=1|\tilde{Z}_L=\tilde{z}_L)$ and $\tilde{T}_{D \to L}=\tilde{t}_L-\tilde{t}_D$ for subjects with $\tilde{X}_L=1$ (see Figure 1(c)). 
Therefore, it holds that
\begin{align}
\tilde{F}_{D \to L}(t|\tilde{Z}_L=\tilde{z}_L,\tilde{X}_L=1)
& = \alpha^{-1}(\tilde{z}_L)\Pr(\tilde{T}_{D \to L} \le t, \tilde{X}_L=1|\tilde{Z}_L=\tilde{z}_L) \nonumber \\
& = \alpha^{-1}(\tilde{z}_L)\Pr(\tilde{t}_{D} \ge \tilde{t}_L - t, \tilde{X}_L=1|\tilde{Z}_L=\tilde{z}_L). \nonumber
\end{align}
Since $\tilde{Z}_L$ contains age at diagnosis and year, $\tilde{t}_L$ is constant conditional of $\tilde{Z}_L$. Then, we have
\begin{align}
&\tilde{F}_{D \to L}(t|\tilde{Z}_L=\tilde{z}_L,\tilde{X}_L=1) \nonumber \\
& = \alpha^{-1}(\tilde{z}_L)\int_{\tilde{t}_L - t}^{\tilde{t}_L}{ \int_{\tilde{t}_L-u}^{\infty}{ \tilde{f}_{\tilde{t}_D,\tilde{T}_{D\to O}}(u,s|\tilde{Z}_L=\tilde{z}_L) ds } du } \nonumber \\
&= \alpha^{-1}(\tilde{z}_L) \sum_{u'=\tilde{t}_L-t}^{\tilde{t}_L-1}{ \int_{u'}^{u'+1}{ \int_{\tilde{t}_L-u}^{\infty}{ \tilde{f}_{\tilde{T}_{D\to O}|\tilde{t}_D}(s|\tilde{t}_D=u, \tilde{Z}_L=\tilde{z}_L) ds } \tilde{f}_{\tilde{t}_D}(u|\tilde{Z}_L=\tilde{z}_L) du } } \nonumber \\
&= \alpha^{-1}(\tilde{z}_L) \sum_{u'=\tilde{t}_L-t}^{\tilde{t}_L-1}{ \int_{u'}^{u'+1}{ \tilde{S}_{D \to O}(\tilde{t}_L-u|\tilde{Z}_D=\tilde{z}_{L-\tilde{t}_L+u'}) \tilde{f}_{\tilde{t}_D}(u|\tilde{Z}_{L-\tilde{t}_L+u'}=\tilde{z}_{L-\tilde{t}_L+u'}) du } }, 
\label{F_DL_int}
\end{align}
by the same as in the expansion of equation (\ref{alpha_int}). 

Again, we use the approximation of $\tilde{S}_{D \to O}(\tilde{t}_L-u|\tilde{Z}_D=\tilde{z}_{L-\tilde{t}_L+u'}) \approx \tilde{S}_{D \to O}(\tilde{t}_L-u'|\tilde{Z}_D=\tilde{z}_{L-\tilde{t}_L+u'})$ on $u \in [u',u'+1)$. 
The equality $\tilde{S}_{D \to O}(\tilde{t}_L-u'|\tilde{Z}_D=\tilde{z}_{L-\tilde{t}_L+u'})=S_O(\tilde{t}_L-u'|Z_D=\tilde{z}_{L-\tilde{t}_L+u'})$ holds from the assumption (C-3). 
Then, coupled with (\ref{int_up_up1}), the equation (\ref{F_DL_int}) leads to
\begin{align}
\tilde{F}_{D \to L}(t|\tilde{Z}_L=\tilde{z}_L,\tilde{X}_L=1) 
&\approx \alpha^{-1}(\tilde{z}_L) \sum_{s=1}^t{S_O(s|Z_D=\tilde{z}_{L-s}) IR(\tilde{z}_{L-s}) \{1-\alpha(\tilde{z}_{L-s}) \}}, \nonumber
\end{align}
for $t=1,2,\cdots$. 
Note that $\tilde{F}_{D \to L}(0|\tilde{Z}_L=\tilde{z}_L,\tilde{X}_L=1)=0$. 
The estimator of $\tilde{F}_{D \to L}(t|\tilde{Z}_L=\tilde{z}_L,\tilde{X}_L=1)$ is defined by
\begin{align}
\hat{F}_{D \to L}(t|\tilde{Z}_L=\tilde{z}_L,\tilde{X}_L=1) 
&= \hat{\alpha}^{-1}(\tilde{z}_L) \sum_{s=1}^t{\hat{S}_O(s|Z_D=\tilde{z}_{L-s}) IR(\tilde{z}_{L-s}) \{1-\hat{\alpha}(\tilde{z}_{L-s}) \}}. \nonumber
\end{align}

\subsection*{B.3. $\tilde{F}_{L \to D}(t|\tilde{Z}_L=\tilde{z}_L,\tilde{X}_L=0)$ in (6)}
\label{use_of_ir_for_F_LD}
It is represented by 
\begin{align}
&\tilde{F}_{L \to D}(t|\tilde{Z}_L=\tilde{z}_L,\tilde{X}_L=0) \nonumber \\
& = 1 - \Pr(\tilde{T}_{L \to D} > t|\tilde{Z}_L=\tilde{z}_L,\tilde{X}_L=0) \nonumber \\
& = 1 - \prod_{s=0}^{t-1}{\Pr(\tilde{t}_{D} > \tilde{t}_L+s+1|\tilde{Z}_L=\tilde{z}_L,\tilde{X}_L=0, \tilde{t}_{D} > \tilde{t}_L+s)} \nonumber \\
& = 1 - \prod_{s=0}^{t-1}{\left\{ 1 - \int_{\tilde{t}_L+s}^{\tilde{t}_L+s+1}{f_{\tilde{t}_D}(u|\tilde{Z}_L=\tilde{z}_L, \tilde{X}_L=0, \tilde{t}_{D} > \tilde{t}_L+s)du} \right\} }, \nonumber 
\end{align}
for $t=1,2,\cdots$. 
Since $\{\tilde{Z}_L=\tilde{z}_L\}$ and $\{\tilde{Z}_{L+s}=\tilde{z}_{L+s}\}$ are the same event and $\{\tilde{X}_L=0, \tilde{t}_D>\tilde{t}_L+s\}$ and $\tilde{X}_{L+s}=0$ are also the same event, $f_{\tilde{t}_D}(u|\tilde{Z}_L=\tilde{z}_L, \tilde{X}_L=0, \tilde{t}_{D} > \tilde{t}_L+s)=f_{\tilde{t}_D}(u|\tilde{Z}_{L+s}=\tilde{z}_{L+s}, \tilde{X}_{L+s}=0)$. 
Then, by the definition (2), 
\begin{align}
&\int_{\tilde{t}_L+s}^{\tilde{t}_L+s+1}{f_{\tilde{t}_D}(u|\tilde{Z}_L=\tilde{z}_L, \tilde{X}_L=0, \tilde{t}_{D} > \tilde{t}_L+s)du} \nonumber \\
& = \int_{\tilde{t}_L+s}^{\tilde{t}_L+s+1}{f_{\tilde{t}_D}(u|\tilde{Z}_{L+s}=\tilde{z}_{L+s}, \tilde{X}_{L+s}=0)du} \nonumber \\
& = IR(\tilde{z}_{L+s}). \nonumber 
\end{align}
Therefore, $\tilde{F}_{L \to D}(t|\tilde{Z}_L=\tilde{z}_L,\tilde{X}_L=0)$ is estimated by
\begin{align}
\hat{F}_{L \to D}(t|\tilde{Z}_L=\tilde{z}_L,\tilde{X}_L=0)
& = 1-\prod_{s=0}^{t-1}{\left\{1-IR(\tilde{z}_{L+s})\right\}}. \nonumber
\end{align}

\section*{C: Log-linear extrapolation of $S_O(t|Z_D)$ beyond the maximum follow-up time}
\label{extrapolate_SO}
We propose a simple extrapolation method based on linear regression for the Kaplan-Meier estimates applied to several time points around the end of follow-up. 
For the subpopulation with $Z_D$, let $\tau_{Z_D}$ be a time point such that on $[0, \tau_{Z_D}]$, $S_O(t|Z_D)$ is well estimated by the Kaplan-Meier estimate $\hat{S}_O(t|Z_D)$. 
We call $\tau_{Z_D}$ the end of follow-up. 
Select several (say $H$) time points near the end of follow-up $\tau_{Z_D}$ and they are denoted by $t^{*}_1, t^{*}_2, \dots, t^{*}_H$ ($t^{*}_1<t^{*}_2< \dots < t^{*}_H$). 
Borrowing the idea of the piecewise exponential distribution to express flexible parametric distribution, we assume that near $\tau_{Z_D}$, the survival function is well approximated by an exponential distribution. 
That is $S_O(t|Z_D) \simeq \exp{(-\lambda t)}$ locally near $\tau_{Z_D}$. 
This motivates to apply a log-linear model $-\log(S_O(t_h^*|Z_D))=\gamma_0+\gamma_1 t_h^*+\epsilon_h$ for $h=1, 2, \cdots, H$ and extrapolate $S_O(t|Z_D)$ by $\exp{(-\hat{\gamma}_0-\hat{\gamma}_1 t)}$ for $t$ after $\tau_{Z_D}$, where $(\hat{\gamma}_0, \hat{\gamma}_1)$ is the least square estimator. 
Denote $\hat{S}_{O}(t^*|Z_D)=\{ \hat{S}_{O}(t_1^*|Z_D), \cdots, \hat{S}_{O}(t_H^*|Z_D) \}^{tr}$. 
The least-squares estimator is calculated by $\hat{\gamma}=A_{H} \{ -\log{ \hat{S}_{O}(t^*|Z_D) }\} $, where
\[ 
A_{H}=
\left\{ 
\left(
\begin{array}{cccc}
1 & 1 & \cdots & 1 \\
t_1^* & t_2^* & \cdots & t_H^* 
\end{array}
\right)
\left(
\begin{array}{cc}
1 & t_1^* \\
\vdots & \vdots \\
1 & t_H^*
\end{array}
\right)
\right\}^{-1}
\left(
\begin{array}{cccc}
1 & 1 & \cdots & 1 \\
t_1^* & t_2^* & \cdots & t_H^* 
\end{array}
\right). 
\]
A similar idea has applied to estimate the mean survival time\cite{Gong2012} and the terminal time points of dose-response curve (Chapter 16 in \citet{Reisfeld2012}). 

By the general large sample theory for the Kaplan-Meier estimator (see chapter 3 in \citet{Fleming1991}), $\hat{S}_{O}(t|Z_D)$ converges in probability to $S_{O}(t|Z_D)$ uniformly in $t \in [0,\tau_{Z_D}]$ as $n_{Z_D}\to \infty$, where $n_{Z_D}$ is the number of subjects with $Z_D$. 
For $t>\tau_{Z_D}$, by using the consistency of $\hat{S}_O(t^*|Z_D)$, we can show easily that $\hat{S}_O(t|Z_D)=\exp{(-\hat{\gamma}_0-\hat{\gamma}_1 t)}$ converges in probability to $S_O(t|Z_D)$ uniformly in $t$. 
Therefore, it can be easily shown that $\hat{S}_{O}(t|Z_D)$ converges in probability to $S_O(t|Z_D)$ uniformly in $t \in [0,\infty)$ as $n_{Z_D}\to \infty$ under the log-linearity assumption. 

\section*{D: Consistency of $S_P(t|Z_D)$}
\label{consistency}
Assume the covariates vector $Z_D$ and $\tilde{Z}_L$ are bounded. 
Let $n_{Z_D}=n_T a_{Z_D}$ where $a_{Z_D} \in (0,1)$ with $\sum_{Z_D}{a_{Z_D}}=1$ and $n_T$ is the total sample size in the cancer registry data. 
We consider limits as $n_T$ approaches infinity. 
Suppose that the approximations of $\alpha(\tilde{z}_L)$ and $\tilde{F}_{D \to L}$ described in Web-Appendices ~B.1 and B.2, respectively, are consistent. 
By the simple algebras and the consistency of $\hat{S}_O(t|Z_D)$, $\hat{\alpha}(\tilde{z}_L)$ converges in probability to $\alpha(\tilde{z}_L)$, and $\hat{ F }_{D\to L}(t|\tilde{Z}_L=\tilde{z}_L,\tilde{X}_L=1)$ converges in probability to $\tilde{F}_{D\to L}(t|\tilde{Z}_L=\tilde{z}_L,\tilde{X}_L=1)$ in each $t=0,1,\cdots$. 
Then, simple algebraic manipulation gives, 
\begin{align}
& \hat{S}_{L\to O}(t|\tilde{Z}_L=\tilde{z}_L,\tilde{X}_L=1) - \tilde{S}_{L\to O}(t|\tilde{Z}_L=\tilde{z}_L,\tilde{X}_L=1) \nonumber \\
& = \int_0^{\tilde{t}_L}{ \left\{\hat{S}_O(t+s|Z_D=\tilde{z}_{L-s}) - S_O(t+s|Z_D=\tilde{z}_{L-s})\right\}d\hat{F}_{D\to L}(s|\tilde{Z}_L=\tilde{z}_L,\tilde{X}_L=1)} \nonumber \\
& - \int_0^{\tilde{t}_L}{ \left\{1-S_O(t+s|Z_D=\tilde{z}_{L-s})\right\} \left\{ d\hat{ F }_{D\to L}(s|\tilde{Z}_L=\tilde{z}_L,\tilde{X}_L=1) - d\tilde{F}_{D\to L}(s|\tilde{Z}_L=\tilde{z}_L,\tilde{X}_L=1) \right\} } \nonumber \\
& \xrightarrow{P} 0, \nonumber
\end{align}
as $n_T \to \infty$ for $t=1,2,\cdots$. 
It follows that $\hat{ S }_{L\to O}(t|\tilde{Z}_L=\tilde{z}_L,\tilde{X}_L=1)$ converges in probability to $\tilde{S}_{L\to O}(t|\tilde{Z}_L=\tilde{z}_L,\tilde{X}_L=1)$ at $t=0,1,\cdots$. 

We prove the consistency of $\hat{S}_P(t|Z_D)$ for $t=1,2,\cdots$ presented in Section 4.4. 
When $t=1$, by the above results, $\hat{S}_P(1|Z_D=\tilde{z}_L)$ presented in Section 4.4 converges in probability to 
\begin{align}
\frac{\tilde{S}_{L\to O}(1|\tilde{Z}_L=\tilde{z}_L) - \alpha(\tilde{z}_L)\tilde{S}_{L\to O}(1|\tilde{Z}_L=\tilde{z}_L, \tilde{X}_L=1)}{\left\{1-\alpha(\tilde{z}_L)\right\}}, 
\label{SP1_limit}
\end{align}
as $n_T \to \infty$. 
Under the assumptions (A-1), (B-1) to (B-3), and (C-1) to (C-3), the equation (\ref{SP1_limit}) is equal to $S_P(1|Z_D=\tilde{z}_L)$. 
It holds that $\hat{S}_P(1|Z_D=\tilde{z}_L)$ converges in probability to $S_P(1|Z_D=\tilde{z}_L)$. 
When $t=2$, since 
\begin{align}
\hat{h}_{\tilde{z}_L}(2,1) &= 1-\frac{\hat{S}_O(2-1|Z_D=\tilde{z}_{L+1})}{\hat{S}_P(2-1|Z_D=\tilde{z}_{L+1})}= 1-\frac{\hat{S}_O(1|Z_D=\tilde{z}_{L+1})}{\hat{S}_P(1|Z_D=\tilde{z}_{L+1})} \nonumber
\end{align}
converges to $1 - \frac{S_O(1|Z_D=\tilde{z}_{L+1})}{S_P(1|Z_D=\tilde{z}_{L+1})}$ and also $\hat{r}(2|\tilde{z}_L)=1-\hat{h}_{\tilde{z}_L}(2,1)\Delta \tilde{F}_1$ converges to $\tilde{S}_{L \to E}(2|\tilde{Z}_L=\tilde{z}_L, \tilde{X}_L=0)$ by the results when $t=1$ and the derivation in Appendix A under the assumptions (A-1), (B-1) to (B-3), and (C-1) to (C-3), 
it also follows that
\begin{align}
\hat{S}_P(2|Z_D=\tilde{z}_L) &= \frac{\tilde{S}_{L\to O}(2|\tilde{Z}_L=\tilde{z}_L) - \hat{ \alpha }(\tilde{z}_L)\hat{ \tilde{S} }_{L\to O}(2|\tilde{Z}_L=\tilde{z}_L, \tilde{X}_L=1)}{\left\{1-\hat{ \alpha }(\tilde{z}_L)\right\} \hat{r}(2|\tilde{z}_L)} \nonumber \\
& \xrightarrow{P} \frac{\tilde{S}_{L\to O}(2|\tilde{Z}_L=\tilde{z}_L) - \alpha(\tilde{z}_L)\tilde{S}_{L\to O}(2|\tilde{Z}_L=\tilde{z}_L, \tilde{X}_L=1)}{\left\{1-\alpha(\tilde{z}_L)\right\} \tilde{S}_{L \to E}(2|\tilde{Z}_L=\tilde{z}_L, \tilde{X}_L=0)} \nonumber \\
& = S_P(2|Z_D=\tilde{z}_L), \nonumber
\end{align}
as $n_T \to \infty$. 
For $t=3$, $\hat{S}_P(3|Z_D=\tilde{z}_L)$ also converges in probability to $S_P(3|Z_D=\tilde{z}_L)$ as $n_T \to \infty$ by using the results of $t=1$ and $2$. 
By repeating this sequentially with $t=4,5,\cdots,K$, we can prove that for each $t=k\ (k=1,2,\cdots K)$, $\hat{S}_P(t|Z_D=\tilde{z}_L)$ is shown to converge in probability to $S_P(t|Z_D=\tilde{z}_L)$ as $n_T\to \infty$.

\end{document}